\documentclass[11pt]{article}

\usepackage[a4paper]{geometry}
\usepackage[utf8]{inputenc} 
\usepackage{lineno}
\usepackage{algorithm}
\usepackage{algorithmicx}
\usepackage{algpseudocode}
\usepackage{graphicx}
\usepackage{caption}
\usepackage{subcaption}
\usepackage{multirow}
\usepackage[dvipsnames]{xcolor} 
  \definecolor{bleu_cite}{RGB}{0,0,255}

\usepackage{amsmath}
\usepackage{bm}
\usepackage{stmaryrd}
\usepackage[utf8]{inputenc}
\usepackage{bbm}
\usepackage{amssymb,amsthm,mathrsfs}
\colorlet{linkequation}{blue}
\colorlet{linkalg}{blue}
\colorlet{linkeq}{blue}
\colorlet{refeq}{green}
\usepackage{etoolbox} 
\usepackage[toc,page]{appendix} 
\usepackage[numbers]{natbib}

\def\eqs{\;}
\newcommand{\map}{\hat{\psi}_i}
\newcommand{\phimap}{\hat{\phi}_i}

\newcommand{\dens}{\texttt{p}_i}

\newcommand{\inter}{\{1, \dots, N\}}
\DeclareMathOperator{\jacob}{J}
\DeclareMathOperator{\hessian}{H}
\newcommand{\exponential}{\operatorname{e}}

\makeatletter
\newcounter {subsubsubsection}[subsubsection]
\renewcommand\thesubsubsubsection{\thesubsubsection .\@alph\c@subsubsubsection}
\newcommand\subsubsubsection{\@startsection{subsubsubsection}{4}{\z@}%
                                     {-3.25ex\@plus -1ex \@minus -.2ex}%
                                     {1.5ex \@plus .2ex}%
                                     {\normalfont\normalsize\bfseries}}
\newcommand*\l@subsubsubsection{\@dottedtocline{3}{10.0em}{4.1em}}
\newcommand*{\subsubsubsectionmark}[1]{}
\makeatother

\DeclareMathOperator*{\argmax}{arg\,max}
\DeclareMathOperator*{\ELBO}{ELBO}

\def\psipop{\psi_{\rm pop}}
\def\iid{\mathop{\sim}_{\rm i.i.d.}}

\newtheorem{lemma}{Independent proposal}
\newtheorem{rem}{Remark}
\newcommand{\trans}[1]{#1^\prime}
\def\thml{\hat{\theta}_{\rm ML}}

\makeatletter
\def\ps@pprintTitle{%
 \let\@oddhead\@empty
 \let\@evenhead\@empty
 \def\@oddfoot{}%
 \let\@evenfoot\@oddfoot}
\makeatother
\title{f-SAEM: A fast Stochastic Approximation of the EM algorithm for nonlinear mixed effects models}
\author{%
\textsc{Belhal Karimi, Marc Lavielle, Eric Moulines}\\
\normalsize  CMAP, Ecole Polytechnique, Universite Paris-Saclay, 91128 Palaiseau, France\\ 
INRIA Saclay, 1 Rue Honoré d'Estienne d'Orves, 91120 Palaiseau, France
}
\date{\today} 

\begin{document}

\maketitle

\begin{abstract}
\noindent The ability to generate samples of the random effects from their conditional distributions  is fundamental for inference in mixed effects models. Random walk Metropolis is widely used  to perform such sampling, but this method is known to converge slowly for medium dimensional problems, or when the joint structure of the distributions to sample is spatially heterogeneous.
The main contribution consists of an independent Metropolis-Hastings (MH) algorithm based on a multidimensional Gaussian proposal that takes into account the joint conditional distribution of the random effects and does not require any tuning. Indeed, this distribution is automatically obtained thanks to a Laplace approximation of the incomplete data model. Such approximation is shown to be equivalent to linearizing the structural model in the case of continuous data. 
Numerical experiments based on simulated and real data illustrate the performance of the proposed methods.
For fitting nonlinear mixed effects models,  the suggested MH algorithm is efficiently combined with a stochastic approximation version of the EM algorithm for maximum likelihood estimation of the global parameters.
\end{abstract}

\section{Introduction}

Mixed effects models are often adopted to take into account the inter-individual variability within a population (see \cite{laviellebook} and the references therein).
Consider a study with $N$ individuals from a same population.
The vector of observations $y_i$ associated to each individual $i$ is assumed to be a realisation of a random variable which depends on a vector of random individual parameters $\psi_i$.
Then, inference on the individual parameter $\psi_i$ amounts to estimate its conditional distribution given the observed data $y_i$.

When the model is a linear (mixed effects) Gaussian model, then this conditional distribution is a normal distribution that can explicitly be computed \cite{verbeke1997linear}. 
For more complex distributions and models, Monte Carlo methods must be used to approximate this conditional distribution.
Most often, direct sampling from this conditional distribution is inefficient and it is necessary to resort to a Markov chain Monte Carlo (MCMC) method for obtaining random samples from this distribution. 
Yet, MCMC requires a tractable likelihood in order to compute the acceptance ratio. When this computation is impossible, a pseudo-marginal Metropolis Hastings (PMMH) has been developed in \cite{andrieu2009pseudo} and consists in replacing the posterior distribution evaluated in the MH acceptance rate by an unbiased approximation.
An extension of the PMMH is the particle MCMC method, introduced in \cite{andrieu2010particle}, where a Sequential Monte Carlo sampler \cite{doucet2000sequential} is used to approximate the intractable likelihood at each iteration. For instance, this method  is relevant when the model is SDE-based (see \cite{donnet2013using}). 
In a fully Bayesian setting, approximation of the posterior of the global parameters can be used to approximate the posterior of the individual parameters using Integrated Nested Laplace Approximation (INLA) introduced in \cite{rue2009approximate}. 
When the goal is to do approximate inference, this method has shown great performances mainly because it approximates each marginal separately as univariate Gaussian distribution. 
In this paper, we focus on developing a method to perform exact inference and do not treat the case of approximate inference algorithms such as the Laplace EM or the First Order Conditional Estimation methods \cite{wang} that can introduce bias in the resulting parameters.

Note that generating random samples from $\dens(\psi_i|y_i;\theta)$ is useful for several tasks to avoid approximation of the model, such as linearisation or Laplace method. 
Such tasks include the estimation of the  population parameters $\theta$ of the model by a maximum likelihood approach, i.e. by maximizing the observed incomplete data likelihood $\texttt{p}(y_1,\ldots y_N;\theta)$ using the Stochastic Approximation of the EM algorithm (SAEM) algorithm combined with a MCMC procedure \cite{kuhn}.
Lastly, sampling from the conditional distributions $\dens(\psi_i|y_i;\theta)$ is also known to be useful for model building. Indeed, in \cite{lavielle2016enhanced}, the authors argue that methods for model assessment and model validation, whether graphical or based on statistical tests, must use samples of the conditional distribution $\dens(\psi_i|y_i;\theta)$ to avoid bias.

Designing a fast mixing sampler for these distributions is therefore  of utmost importance to perform Maximum Likelihood Estimation (MLE) using the SAEM algorithm. The most common MCMC method for nonlinear mixed effects (NLME) models is the \emph{random walk Metropolis} (RWM) algorithm \cite{mh:robert,robertsoptimal, laviellebook}. This method is implemented in software tools such as Monolix, NONMEM, the saemix R package \cite{comets} and the nlmefitsa Matlab function. 
Despite its simplicity, it has been successfully used in many classical examples of pharmacometrics. Nevertheless, it can show its limitations when the dependency structure of the individual parameters is complex. 
Yet, maintaining an optimal acceptance rate (advocated in \cite{roberts}) most often implies very small moves and therefore a very large number of iterations in medium and high dimensions since no information of the geometry of the target distribution is used.

The Metropolis-adjusted Langevin algorithm (MALA)
 uses evaluations of the gradient of the target density for proposing new states
which are accepted or rejected using the Metropolis-Hastings algorithm \cite{robertsmala, stramer}. 
Hamiltonian Monte Carlo (HMC) is another MCMC  algorithm that exploits information about the geometry of the target distribution in order to efficiently explore the space by selecting transitions that can follow  contours of high probability mass  \cite{betancourt2017conceptual}.
The No-U-Turn Sampler (NUTS) is an extension to HMC that allows an automatic and optimal selection of some of the settings required by the algorithm, \cite{brooks2011handbook,hoffman2014no}.
Nevertheless, these methods may be difficult to use in practice, and are computationally involved, in particular when the structural model is a complex ODE based model.
The algorithm we propose is an independent Metropolis-Hastings (IMH) algorithm, but for which the proposal is a Gaussian approximation of the target distribution.
For general data model (i.e. categorical, count or time-to-event data models or continuous data models), the Laplace approximation of the incomplete pdf $\dens(y_i;\theta)$ leads to  a Gaussian approximation of the conditional distribution $\dens(\psi_i |y_i;\theta)$.

In the special case of continuous data, linearisation of the model leads, by definition, to a Gaussian linear model for which the conditional distribution of the individual parameter $\psi_i$ given the data $y_i$ is a multidimensional normal distribution that can be computed.
Therefore, we design an independent sampler using this multivariate Gaussian distribution to sample from the target conditional distribution and embed this procedure in an exact inference algorithm, the SAEM, to speed the convergence of the vector of estimations of the global parameters $\hat{\theta}$.

The paper is organised as follows. Mixed effects models for continuous and noncontinuous data are presented in Section~2. 
The standard MH for NLME models is described in Section~3.
The proposed method, called the nlme-IMH, is introduced in Section~4 as well as the f-SAEM, a combination of this new method with the SAEM algorithm for estimating the population parameters of the model. 
Numerical examples illustrate, in Section~5, the practical performances of the proposed method, both on a continuous pharmacokinetics (PK) model and a time-to-event example. A Monte Carlo study confirms that this new SAEM algorithm shows a faster convergence to the maximum likelihood estimate.

\section{Mixed Effect Models}

\subsection{Population approach and hierarchical models}

In the sequel, we adopt a population approach, where we consider $N$ individuals and $n_i$ observations per individual $i$. The set of observed data is $y = (y_i, 1\leq i \leq N)$ where $y_i = (y_{ij}, 1\leq j \leq n_i)$ are the observations for individual $i$. For the sake of clarity, we assume that each observation $y_{ij}$ takes its values in some subset of $\mathbb{R}$. The distribution of the $n_i-$vector of observations $y_i$ depends on a vector of individual parameters $\psi_i$ that takes its values in a subset of $\mathbb{R}^{p}$.

We assume that the pairs $(y_i,\psi_i)$ are mutually independent and consider a parametric framework: the joint distribution of $(y_i,\psi_i)$ is denoted by $\dens(y_i,\psi_i;\theta)$, where $\theta$ is the vector of parameters of the model. 
A natural decomposition of this joint distribution reads
\begin{equation}
    \dens(y_i,\psi_i;\theta) = \dens(y_i|\psi_i;\theta)\dens(\psi_i;\theta)\eqs,
\end{equation}
where $\dens(y_i|\psi_i;\theta)$ is the conditional distribution of the observations given the individual parameters, and where $\dens(\psi_i;\theta)$ is the so-called population distribution used to describe the distribution of the individual parameters within the population.

A particular case of this general framework consists in describing each individual parameter $\psi_i$ as the sum of a typical value $\psipop$ and a vector of individual random effects $\eta_i$:
\begin{equation} \label{indiv_param1}
\psi_i = \psipop+\eta_i\eqs.
\end{equation}
In the sequel, we assume that the random effects are distributed according to a multivariate Gaussian distribution: $\eta_i \iid \mathcal{N}(0,\Omega)$. Extensions of this general model are detailed in Appendix~\ref{appendix:modelextensions}.

\subsection{Continuous data models}  \label{sec:cont}

A regression model is used to express the link between continuous observations and individual parameters:
\begin{equation}\label{continuousmodel}
y_{ij} = f_i(t_{ij},\psi_i) + \varepsilon_{ij}\eqs,
\end{equation}
where $y_{ij}$ is the $j$-th observation for individual $i$ measured at index $t_{ij}$, $\varepsilon_{ij}$ is the residual error. It is assumed that for any index $t$, $\psi \to f_i(t,\psi)$ is twice differentiable in $\psi$.

We start by assuming that the residual errors are independent and normally distributed with zero-mean and a constant variance $\sigma^2$. Let $t_i=(t_{ij}, 1\leq n_i)$ be the vector of observation indices for individual $i$. Then, the model for the observations reads:
\begin{equation}\label{eq:vectorobs}
 y_i|\psi_i \sim \mathcal{N}(f_i(\psi_i),\sigma^2\texttt{Id}_{n_i\times n_i}) \quad \textrm{where} \quad f_i(\psi_i) = (f_i(t_{i,1},\psi_i), \ldots , f_i(t_{i,n_i},\psi_i)) \eqs.
\end{equation}
If we assume that $\psi_i \iid \mathcal{N}(\psipop,\Omega) $, then the parameters of the model are $\theta= (\psipop, \Omega, \sigma^2)$.

\begin{rem}
An extension of this model consists in assuming that the variance of the residual errors is not constant over time, i.e., $\varepsilon_{ij} \sim \mathcal{N}(0,g(t_{ij}, \psi_i)^2)$.
Such extension includes proportional error models ($g=bf$) and combined error models ($g=a+bf$) \cite{laviellebook} but the proposed method remains the same whatever the residual error model is.
\end{rem}

 \subsection{Noncontinuous data models}\label{noncontinuousdef}
Noncontinuous data models include categorical data models \cite{savic, agresti}, time-to-event data models \cite{mbogning, andersen}, or count data models \cite{savic}.
A categorical outcome $y_{ij}$ takes its value in a set $\{1, \dots, L\}$ of $L$ categories. Then, the model is defined by the conditional probabilities $\left(\mathbb{P}(y_{ij} = \ell | \psi_i),1 \leq \ell \leq L\right)$, that depend on the vector of individual parameters $\psi_i$ and may be a function of the time $t_{ij}$.

In a time-to-event data model, the observations are the times at which events occur. An event may be one-off (e.g., death, hardware failure) or repeated (e.g., epileptic seizures, mechanical incidents).
To begin with, we consider a model for a one-off event. The survival function $S(t)$ gives the probability that the event happens after time $t$:
\begin{equation}\label{survival}
S(t)  \triangleq \mathbb{P}(T >t) = \exp\left\{-\int_{0}^{t}h(u)\textrm{d}u\right\}\eqs,
\end{equation}
where $h$ is called the hazard function. 
In a population approach, we consider a parametric and individual hazard function $h(\cdot,\psi_i)$.
The random variable representing the time-to-event for individual $i$ is typically written $T_i$ and may possibly be right-censored. Then, the observation $y_i$ for individual $i$ is
\begin{equation}
    y_i = \left\{
    \begin{array}{ll}
        T_i & \mbox{if } T_i \leq \tau_c \\
        "T_i > \tau_c" & \mbox{otherwise}\eqs,
    \end{array}
\right.
\end{equation}
where $\tau_c$ is the censoring time and $"T_i > \tau_c"$ is the information that the event occurred after the censoring time. 

For repeated event models, times when events occur for individual $i$ are random times $(T_{ij}, 1\leq j \leq n_i)$  for which conditional survival functions can be defined:
\begin{equation}
\mathbb{P}(T_{ij} > t|T_{i(j-1)}=t_{i(j-1)}) = \exp\left\{-\int_{t_{i(j-1)}}^{t}h(u, \psi_i)\textrm{d}u\right\}\eqs.
\end{equation}
Here, $t_{ij}$ is the observed value of the random time $T_{ij}$.
If the last event is right censored, then the last observation $y_{i,n_i}$ for individual $i$ is the information that the censoring time has been reached $"T_{i,n_i} > \tau_c"$. The conditional pdf of $y_i = (y_{ij},1\leq n_i)$ reads (see \cite{laviellebook} for more details)
\begin{equation} \label{pdf_tte}
\dens(y_i |\psi_i) = \exp\left\{-\int_{0}^{\tau_c}h(u, \psi_i)\textrm{d}u \right\} \prod_{j=1}^{n_i-1} h(t_{ij},\psi_i)\eqs.
\end{equation}

\section{Sampling from conditional distributions}

\subsection{The conditional distribution of the individual parameters}
Once the conditional distribution of the observations $\dens(y_i|\psi_i ; \theta)$ and the marginal distribution of the individual parameters $\psi_i$ are defined, the joint distribution $\dens(y_i,\psi_i ; \theta)$ and the conditional distribution $\dens(\psi_i | y_i ; \theta)$ are implicitly specified.
This conditional distribution $\dens(\psi_i | y_i ; \theta)$ plays a crucial role for inference in NLME models.

One of the main task is to compute the maximum likelihood (ML) estimate of $\theta$
\begin{equation} \label{ml}
\thml = \arg \max \limits_{\theta \in \mathbb{R}^d} {\cal L}(\theta , y)\eqs,
\end{equation}
where ${\cal L}(\theta , y) = \log \texttt{p}(y;\theta)$.
In NLME models, this optimization is solved by using a surrogate function defined as the conditional expectation of the complete data log-likelihood \cite{mclachlan2007algorithm}. 
The SAEM is an iterative procedure for ML estimation that requires to generate one or several samples from this conditional distribution at each iteration of the algorithm.
Once the ML estimate $\thml$ has been computed, the observed Fisher information matrix noted $I(\thml, y) = - \nabla_\theta^2 {\cal L}(\thml , y)$ can be derived thanks to the Louis formula \cite{louis} which expresses $I(\thml, y)$ in terms of the conditional expectation and covariance of the complete data log-likelihood.
Such procedure also requires to sample from the conditional distributions $\dens(\psi_i | y_i ; \thml)$ for all $i \in \inter$.

Samples from the conditional distributions might also be used to define several statistical tests and diagnostic plots for models assessment. 
It is advocated in \cite{lavielle2016enhanced} that such samples should be preferred to the modes of these distributions (also called \emph{Empirical Bayes Estimate}(EBE), or \emph{Maximum a Posteriori Estimate}), in order to provide unbiased tests and plots. For instance, a strong bias can be observed when the EBEs are used for testing the
distribution of the parameters or the correlation between random effects.

In short, being able to sample individual parameters from their conditional distribution is essential in nonlinear mixed models. It is therefore necessary to design an efficient method to sample from this distribution.

\subsection{The Metropolis-Hastings Algorithm}\label{sec:rwm}

Metropolis-Hasting (MH) algorithm is a powerful MCMC procedure widely used for sampling from a complex distribution \cite{brooks2011handbook}.
To simplify the notations, we remove the dependency on $\theta$.
For a given individual $i \in \inter$, the MH algorithm, to sample from the conditional distribution $\dens(\psi_i | y_i)$, is described as:
\begin{algorithm}[H]
\textbf{Initialization}: Initialize the chain sampling $\psi_i^{(0)}$ from some initial distribution $\xi_i$ .\\
\textbf{Iteration k}: given the current state of the chain $\psi_i^{(k-1)}$:
\begin{enumerate}
\item Sample a candidate $\psi_i^c$ from a proposal distribution $q_i( \, \cdot \, | \psi_i^{(k-1)})$.
\item Compute the MH ratio:
\begin{equation}\label{eq:MHratio}
\alpha(\psi_i^{(k-1)},\psi_i^{c}) = 
\frac{\dens(\psi_i^{c}|y_i)}{\dens(\psi_i^{(k-1)}|y_i)}
\frac{q_i(\psi_i^{(k-1)}|\psi_i^c)}{q_i(\psi_i^{c}|\psi_i^{(k-1)}) }\eqs.
\end{equation}
\item Set $\psi_i^{(k)}=\psi_i^c$ with probability $\min (1,\alpha(\psi_i^{c},\psi_i^{(k-1)})$ (otherwise, keep $\psi_i^{(k)}=\psi_i^{(k-1)}$).
\end{enumerate}
\caption{Metropolis-Hastings algorithm}
\label{alg:mh}
\end{algorithm}

Under weak conditions, $(\psi_i^{(k)}, k\geq 0)$ is an ergodic Markov chain whose distribution converges to the target $\dens(\psi_i | y_i)$ \cite{brooks2011handbook}.

Current implementations of the SAEM algorithm in Monolix \cite{chan}, saemix (R package) \cite{comets}, nlmefitsa (Matlab) and NONMEM \cite{beal} mainly use the same combination of proposals. 
The first proposal is an independent MH algorithm which consists in sampling the candidate state directly from the prior distribution of the individual parameter $\psi_i$. The MH ratio then reduces to $\dens(y_i|\psi_i^{c})/\dens(y_i|\psi_i^{(k)})$ for this proposal. 

The other proposals are component-wise and block-wise  random walk procedures \cite{metropolis} that updates different components of $\psi_i$ using univariate and multivariate Gaussian proposal distributions. 
These proposals are centered at the current state with a diagonal variance-covariance matrix; the variance terms are adaptively adjusted at each iteration in order to reach some target acceptance rate \cite{atchade,laviellebook}.
Nevertheless, those proposals fail to take into account the nonlinear dependence structure of the individual parameters.
 
A way to alleviate these problems is to use a proposal distribution derived from a discretised Langevin diffusion whose drift term is the gradient of the logarithm of the target density leading to the Metropolis Adjusted Langevin Algorithm (MALA).
The MALA proposal is a multivariate Gaussian with the following mean $\mu_{i, {\rm MALA}}^{(k)}$ and covariance matrix $\Gamma_ {\rm MALA}$:
\begin{equation}\label{eq:update.mala}
\mu_{i, {\rm MALA}}^{(k)} = \psi_i^{(k)} + \gamma \nabla_{\psi_i} \log \dens (\psi_i^{(k)}|y_i) \quad \textrm{and} \quad \Gamma_ {\rm MALA} = 2\gamma \mathsf{I}_p
\end{equation}
where $\gamma$ is a positive stepsize and $\mathsf{I}_p$ is the identity matrix in $\mathbb{R}^{p \times p}$. 
These methods appear to behave well for complex models but still do not take into consideration the multidimensional structure of the individual parameters. 
Recent works include efforts in that direction, such as the Anisotropic MALA for which the covariance matrix of the proposal depends on the gradient of the target measure \cite{kuhnamala}, the Tamed Unadjusted Langevin Algorithm \cite{brosse2017tamed} based on the coordinate-wise taming of superlinear drift coefficients and a multidimensional extension of the Adaptative Metropolis algorithm \cite{haario2001adaptive} simultaneously estimating the covariance of the target measure and coercing the acceptance rate, see \cite{vihola2012robust}.

The MALA algorithm is a special instance of the Hybrid Monte Carlo (HMC), introduced in \cite{neal2011mcmc}; see \cite{brooks2011handbook} and the references therein, and consists in augmenting the state space with an auxiliary variable $p$, known as the velocity in Hamiltonian dynamics.
This algorithm belongs to the class of data augmentation methods. Indeed, the potential energy is augmented with a kinetic energy, function of an added auxiliary variable. The MCMC procedure then consists in sampling from this augmented posterior distribution.
All those methods aim at finding the proposal $q$ that accelerates the convergence of the chain. Unfortunately they are computationally involved (even in small and medium dimension settings, the computation of the gradient or the Hessian can be overwhelming) and can be difficult to implement (stepsizes and numerical derivatives need to be tuned and implemented).

We see in the next section how to define a multivariate Gaussian proposal for both continuous and noncontinuous data models, that is easy to implement and that takes into account the multidimensional structure of the individual parameters in order to accelerate the MCMC procedure.

\section{The nlme-IMH and the f-SAEM}
In this section, we assume that the individual parameters $(\psi_1, \dots, \psi_N)$ are independent and normally distributed with mean $(m_1,\dots, m_N)$ and covariance $\Omega$. 
The MAP estimate, for individual $i$, is the value of $\psi_i$ that maximizes the conditional distribution $\dens(\psi_i|y_i,\theta)$: 
\begin{equation}\label{mapdef}
\map = \arg \max \limits_{\psi_i \in \mathbb{R}^p} \dens(\psi_i|y_i) = \arg \max \limits_{\psi_i \in \mathbb{R}^p} \dens(y_i|\psi_i)\dens(\psi_i) \eqs.
\end{equation}
\subsection{Proposal based on Laplace approximation}
For both continuous and noncontinuous data models, the goal is to find a simple proposal, a multivariate Gaussian distribution in our case, that approximates the target distribution $\dens(\psi_i|y_i)$.
For general MCMC samplers, it is shown in \cite{roberts2011quantitative} that the mixing rate in total variation depends on the expectation of the acceptance ratio under the proposal distribution which is also directly related to the ratio of the proposal to the target in the special case of independent samplers (see \cite{mengersen,roberts2011quantitative}).
This observation naturally suggests to find a proposal which approximates the target.
\cite{freitas} advocates the use a multivariate Gaussian distribution whose parameters are obtained by minimizing the Kullback-Leibler divergence between a multivariate Gaussian variational candidate distribution and the target distribution.
In \cite{andrieu2008tutorial} and the references therein, an adaptative Metropolis algorithm is studied and reconciled to a KL divergence minimisation problem where the resulting multivariate Gaussian distribution can be used as a proposal in a IMH algorithm. Authors note that although this proposal might be a sensible choice when it approximates well the target, it can fail when the parametric form of the proposal is not sufficiently rich. 
Thus, other parametric forms can be considered and it is suggested in \cite{andrieu2006ergodicity} to consider mixtures, finite or infinite, of distributions belonging to the exponential family.

In general, this optimization step is difficult and computationally expensive since it requires to approximate (using Monte Carlo integration for instance) the integral of the log-likelihood with respect to the variational candidate distribution.

\begin{lemma}\label{lemma1}
We suggest a Laplace approximation of this conditional distribution as described in \cite{rue2009approximate} which is the multivariate Gaussian distribution with mean $\map$ and variance-covariance

\begin{equation}\label{noncont:covariance}
 \Gamma_i = \left(- \hessian_{\map}+\Omega^{-1}\right)^{-1}\eqs,
\end{equation}
where $\hessian_{\map} \in \mathbb{R}^{p \times p}$ is the Hessian of $\log\left(\dens(y_i|\psi_i)\right)$ evaluated at $\map$.
\end{lemma}
Mathematical details for computing this proposal are postponed to Appendix \ref{appendix:noncontinuous}.
We use this multivariate Gaussian distribution  as a proposal in our IMH algorithm introduced in the next section, for both continuous and noncontinuous data models.

\begin{rem}
Note that the resulting proposal distribution is based on the assumption that, in model \eqref{indiv_param1}, the random effects $\eta_i$ are normally distributed. When this assumption does not hold, our method exploits the same Gaussian proposal, where the variance $\Omega$ in \eqref{noncont:covariance} is calculated explicitely.
Consider the following example: the random effects $\eta_i$ in \eqref{indiv_param1} are no longer distributed according to a multivariate Gaussian distribution but a multivariate Student distribution with $d$ degrees of freedom, zero mean and a prior shape matrix $\xi$ such that $\eta_i \sim t_d(0, \xi)$. 
Then the vector of parameters of the model is $\theta= ( \psi_{\rm pop}, \Omega, \sigma^2)$ where $\Omega = \frac{d}{d - 2}\xi$ is the prior covariance matrix.
In that case, our method uses the Independent proposal 1 and computes the MH acceptance ratio \eqref{eq:MHratio}  with the corresponding multivariate Student density $\dens(\psi_i)$.
\end{rem}

We shall now see another method to derive a Gaussian proposal distribution in the specific case of continuous data models (see \eqref{continuousmodel}).

\subsection{Nonlinear continuous data models} \label{mh:nonlinear}
When the model is described by \eqref{continuousmodel}, the approximation of the target distribution can be done twofold: either by using the Laplace approximation, as explained above, or by linearizing the structural model $f_i$ for any individual $i$ of the population.
using \eqref{continuousmodel} and \eqref{mapdef}, the MAP estimate can thus be derived as: 
\begin{equation}\label{mapdef2}
\map = \arg \min \limits_{\psi_i \in \mathbb{R}^p} \left( \frac{1}{\sigma^2} \| y_{i} - f_i(\psi_i) \|^2 + \trans{(\psi_i-m_i)}\Omega^{-1}(\psi_i-m_i) \right)\eqs.
\end{equation}
where $f_i(\psi_i)$ is defined by \eqref{eq:vectorobs} and $\trans{A}$ is the transpose of the matrix $A$.

We linearize the structural model $f_i$ around the MAP estimate $\map$:
\begin{equation}
f_i(\psi_i) \approx f_i(\map) + \jacob_{f_i(\map)}(\psi_i - \map)\eqs,
\end{equation}
where $\jacob_{f_i(\map)} \in \mathbb{R}^{n_i \times p}$ is the Jacobian of $f_i$ evaluated at $\map$. 
Defining $z_i := y_i - f_i(\map) + \jacob_{f_i(\map)} \map$, this expansion yields the following linear model:
\begin{equation}\label{linear}
z_i = \jacob_{f_i(\map)} \psi_i + \varepsilon_i\eqs.
\end{equation}

We can directly use the definition of the conditional distribution under a linear model (see \eqref{map_lin} in Appendix \ref{appendix:linearmodel}) to get an expression of the conditional covariance $\Gamma_i$ of $\psi_i$ given $z_i$ under \eqref{linear}:
\begin{equation}\label{cont:proposalcov}
\Gamma_i =\left(\frac{ \trans{\jacob_{f_i(\map)}} \jacob_{f_i(\map)} }{\sigma^2} + \Omega^{-1}\right)^{-1}\eqs.
\end{equation}
Using \eqref{mapdef2} and the definition of the conditional distribution under a linear model we obtain that $\mu_i = \map$ (See Appendix \ref{appendix:conditionalmode} for details).
We note that the mode of the conditional distribution of $\psi_i$ in the nonlinear model \eqref{continuousmodel} is also the mode and the mean of the conditional distribution of $\psi_i$ in the linear model \eqref{linear}.

\begin{lemma} \label{lemma:cont}
In the case of continuous data models, we propose to use the multivariate Gaussian distribution, with mean $\map$ and variance-covariance matrix $\Gamma_i$ defined by \eqref{cont:proposalcov} as a proposal for an independent MH algorithm avoiding the computation of an Hessian matrix.
\end{lemma}
We can note that linearizing the structural model is equivalent to using the Laplace approximation with the expected information matrix. Indeed:
\begin{equation}\label{eq:expectedfim}
 \mathbb{E}_{y_i|\map}\left(-\hessian_{l(\map)} \right) =  \frac{\trans{\jacob_{f_i(\map)}} \jacob_{f_i(\map)} }{\sigma^2}\eqs.
\end{equation}

\begin{rem}
When the model is linear, the probability of accepting a candidate generated with this proposal is equal to 1.
\end{rem}
\begin{rem}If we consider a more general error model, $\varepsilon_i \sim \mathcal{N}(0,\Sigma(t_i,\psi_i))$ that may depend on the individual parameters $\psi_i$ and the observation times $t_i$, then the conditional variance-covariance matrix reads:
\begin{equation}
 \Gamma_i = \left( \trans{\jacob_{f_i(\map)}} \Sigma(t_i,\map)^{-1} \jacob_{f_i(\map)} + \Omega^{-1}\right)^{-1}\eqs.
\end{equation}
\end{rem}
\begin{rem}
In the model \eqref{indiv_param2}, the transformed variable $\phi_i = u(\psi_i)$ follows a normal distribution. Then a candidate  $\phi_i^c$ is drawn from the multivariate Gaussian proposal with parameters:
\begin{align}
& \mu_i = \phimap\eqs,\\
& \Gamma_i = \left( \frac{\trans{\jacob_{f_i(u^{-1}(\phimap))}}\jacob_{f_i(u^{-1}(\phimap))}}{\sigma^2} + \Omega^{-1}\right)^{-1}\eqs,
\end{align}
where $\phimap = \arg \max \limits_{\phi_i \in \mathbb{R}^p} \dens(\phi_i|y_i)$ and finally the candidate vector of individual parameters is set to $\psi_i^c = u^{-1}(\phi_i^c)$
\end{rem}

%
%
%
%
%
%

These approximations of the conditional distribution $\dens(\psi_i|y_i)$ lead to our nlme-IMH algorithm, an Independent Metropolis-Hastings (IMH) algorithm for NLME models. For all individuals $i \in \inter$, the algorithm is defined as:
\begin{algorithm}[H]
\textbf{Initialization}: Initialize the chain sampling $\psi_i^{(0)}$ from some initial distribution $\xi_i$ .\\
\textbf{Iteration t}: Given the current state of the chain $\psi_i^{(t-1)}$:
\begin{enumerate}
\item Compute the MAP estimate:
\begin{equation}
\map^{(t)} = \arg \max \limits_{\psi_i \in \mathbb{R}^p} \dens(\psi_i|y_i)\eqs.
\end{equation}
\item Compute the covariance matrix $\Gamma_i^{(t)}$ using either \eqref{noncont:covariance} or \eqref{cont:proposalcov}. 
\item Sample a candidate $\psi_i^c$ from a the independent proposal $\mathcal{N}(\map^{(t)},\Gamma_i^{(t)})$ denoted $q_i(\cdot|\map^{(t)})$.
\item Compute the MH ratio:
\begin{equation}\label{eq:acceptancerate}
\alpha(\psi_i^{(t-1)},\psi_i^{c}) = 
\frac{\dens(\psi_i^{c}|y_i)}{\dens(\psi_i^{(t-1)}|y_i)}
\frac{q_i(\map^{(t)} |\psi_i^c)}{q_i(\psi_i^{c}|\map^{(t)} )} \eqs.
\end{equation}
\item Set $\psi_i^{(t)}=\psi_i^c$ with probability $\min (1,\alpha(\psi_i^{c},\psi_i^{(t-1)}))$ (otherwise, keep $\psi_i^{(t)}=\psi_i^{(t-1)}$).
\end{enumerate}
\caption{The nlme-IMH algorithm}
\label{alg:imh}
\end{algorithm}

This method shares some similiarities with \cite{titsias2018agrad} that suggests to perform a Taylor expansion of $\dens(y_i|\psi_i)$ around the current state of the chain, leaving $\dens(\psi_i)$  unchanged.

\begin{rem}
Although a multivariate Gaussian proposal is used in our presentation of the nlme-IMH, other type of distributions could be adopted. For instance, when the target distribution presents heavy tails, a Student distribution with a well-chosen degree of freedom could improve the performance of the independent sampler. In such case, the parameters of the Gaussian proposal are used to shift and scale the Student proposal distribution and the acceptance rate \eqref{eq:acceptancerate} needs to be modified accordingly.
The numerical applications in Section~5 are performed using a Gaussian proposal but comparisons with a Student proposal distribution are given in Appendix~\ref{appendix:numerical.warfa}.
\end{rem}

\subsection{Maximum Likelihood Estimation}\label{sec:saem_alg}
The ML estimator defined by \eqref{ml} is computed using the Stochastic Approximation of the EM algorithm (SAEM) \cite{lavielle}.
The SAEM algorithm is described as follows:
\begin{algorithm}[H]
\textbf{Initialization}: $\theta_0$, an initial parameter estimate and $M$, the number of MCMC iterations.\\
\textbf{Iteration k}: given the current model parameter estimate $\theta_{k-1}$:
\begin{enumerate}

\item {\bf Simulation step:} for $i \in \inter$, draw vectors of individual parameters $(\psi_1^{(k)}, \dots, \psi_N^{(k)})$ after $M$ transitions of Markov kernels \\ $(\Pi_1^{(k)}(\psi_1^{(k-1)}\,,\,\cdot), \dots,(\Pi_N^{(k)}(\psi_N^{(k-1)}\,,\,\cdot))$ which admit as unique limiting distributions the conditional distributions $(\texttt{p}_1(\psi_1|y_1;\theta_{k-1}), \dots, \texttt{p}_N(\psi_N|y_N;\theta_{k-1}))$,
\item {\bf Stochastic approximation step:} update the approximation of the conditional expectation $\mathbb{E}\left[\log \texttt{p}(y,\psi; \theta) | y, \theta_{k-1}\right]$:
\begin{equation}
Q_k(\theta) = Q_{k-1}(\theta) + \gamma_k\left(\sum_{i=1}^{N}{\log \dens(y_i,\psi_i^{(k)}; \theta)} - Q_{k-1}(\theta)\right)\eqs,
\end{equation}

where $\{\gamma_k\}_{k>0}$ is a sequence of decreasing stepsizes with $\gamma_1 = 1$.
\item {\bf Maximisation step:} Update the model parameter estimate:
\begin{equation}
\theta_k = \arg \max \limits_{\theta \in \mathbb{R}^d} Q_k(\theta)\eqs.
\end{equation}
\end{enumerate}
\caption{The SAEM algorithm}
\label{alg:SAEM}
\end{algorithm}

The SAEM algorithm is implemented in most sofware tools for NLME models and its convergence is studied in \cite{lavielle, kuhn,kuhnamala}.
The practical performances of SAEM are closely linked to the settings of SAEM.
In particular, the choice of the transition kernel $\Pi$ plays a key role. 
The transition kernel $\Pi$ is directly defined by the proposal(s) used for the MH algorithm.

We propose a fast version of the SAEM algorithm using our resulting independent proposal distribution called the f-SAEM. 
The simulation step of the f-SAEM is achieved using the nlme-IMH algorithm (see algorithm \ref{alg:imh}) for all individuals $i \in \inter$ and the next steps remain unchanged.
In practice, the number of transitions $M$ is small since the convergence of the SAEM does not require the convergence of the MCMC at each iteration \cite{kuhn}.
In the sequel, we carry out numerous numerical experiments to compare our nlme-IMH algorithm to state-of-the-art samplers and assess its relevance in a MLE algorithm such as the SAEM.

\section{Numerical Examples}\label{sec:numericalexamples}

\subsection{A pharmacokinetic example}
\subsubsection{Data and model}

\begin{figure}[thp]
\begin{center}
\includegraphics[width=\textwidth]{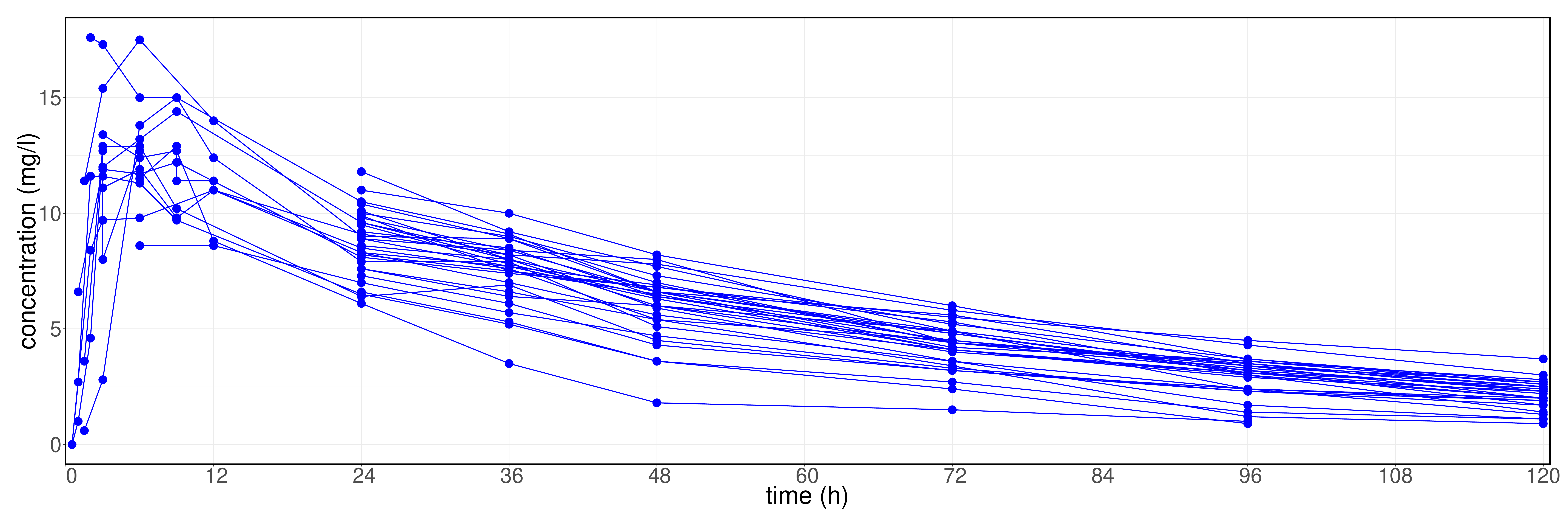}
\caption{Warfarin concentration (mg/l) over time (h) for 32 subjects}
\label{warfa_data}
\end{center}
\end{figure}

32 healthy volunteers received a 1.5 mg/kg single oral dose
of warfarin, an anticoagulant normally used in the prevention of thrombosis \cite{oreilly}.
Figure~\ref{warfa_data} shows the warfarin plasmatic concentration measured at different times for these patients (the single dose was given at time 0 for all the patients).

We consider a one-compartment pharmacokinetics (PK) model for oral administration, assuming first-order absorption and linear elimination processes:
\begin{equation} \label{pkmodel}
f(t,ka, V, k) = \frac{D\,ka}{V(ka - k)}(\exponential^{-ka\,t}-\exponential^{-k\,t})\eqs,
\end{equation}
where $ka$ is the absorption rate constant, $V$ the volume of distribution , $k$ the elimination rate constant, and $D$ the dose of drug administered.
Here, $ka$, $V$ and $k$ are  PK parameters that can change from one individual to another.
Let $ \psi_i=(ka_i, V_i, k_i)$ be the vector of individual PK parameters for individual $i$.
The model for the $j$-th measured concentration, noted $y_{ij}$, for individual $i$ writes:
\begin{equation}
y_{ij} = f(t_{ij},\psi_i)+ \varepsilon_{ij}\eqs.
\end{equation}

We assume in this example that the residual errors are independent and normally distributed with mean 0 and variance $\sigma^2$.
Lognormal distributions are used for the three PK parameters:
\begin{align}
 \log(ka_i) \sim \mathcal{N}(\log(ka_{\rm pop}), \omega^2_{ka})\eqs, \log(V_i) \sim \mathcal{N}(\log(V_{\rm pop}), \omega^2_{V})\eqs,
 \log(k_i) \sim \mathcal{N}(\log(k_{\rm pop}), \omega^2_{k})\eqs.
\end{align}

This is a specific instance of the nonlinear mixed effects model for continuous data described in Section~\ref{sec:cont}. We thus use the multivariate Gaussian proposal whose mean and covariance are defined by \eqref{cont:proposalmean} and \eqref{cont:proposalcov}.
In such case the gradient can be explicitly computed. 
Nevertheless, for the method to be easily extended to any structural model, the gradient is calculated using Automatic Differentiation \cite{griewank2008evaluating} implemented in the R package ``Madness'' \cite{pav2016madness}. 

\subsubsection{MCMC Convergence Diagnostic}\label{sec:mcmc_continuous}
We study in this section the behaviour of the MH algorithm used to sample individual parameters from the conditional distribution $\dens(\psi_i | y_i ; \theta)$.
We consider only one of the 32 individuals for this study and fix $\theta$ close to the ML estimate obtained with the SAEM algorithm, implemented in the saemix R package \cite{comets}:
$ka_{\rm pop} =1$, $V_{\rm pop}= 8$, $k_{\rm pop}=0.01$,  $\omega_{ka}=0.5$, $\omega_{V}=0.2$, $\omega_{k}=0.3$ and $\sigma^2=0.5$.

We run the classical version of MH implemented in the saemix package and for which different transition kernels are used successively at each iteration:
independent proposals from the marginal distribution $\dens(\psi_i)$,
component-wise random walk and block-wise random walk. We compare it to our proposed algorithm \ref{alg:imh}.

\begin{figure}[ht]
  \centering
  \begin{subfigure}{\linewidth}
    \centering
    \includegraphics[width=.8\linewidth]{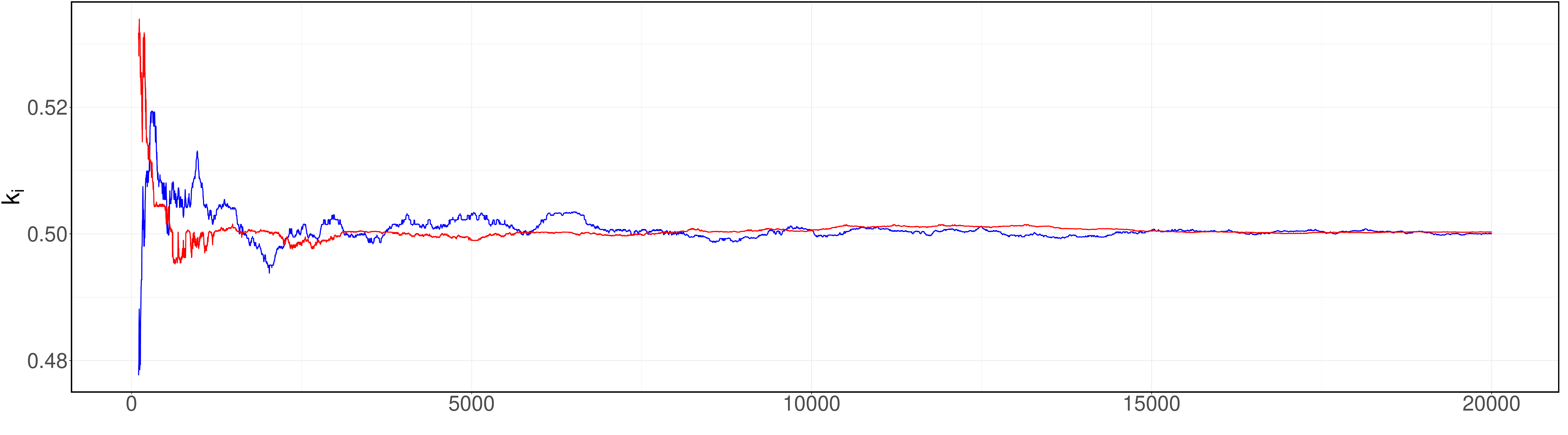}
  \end{subfigure}

  \begin{subfigure}{\linewidth}
    \centering
    \includegraphics[width=.8\linewidth]{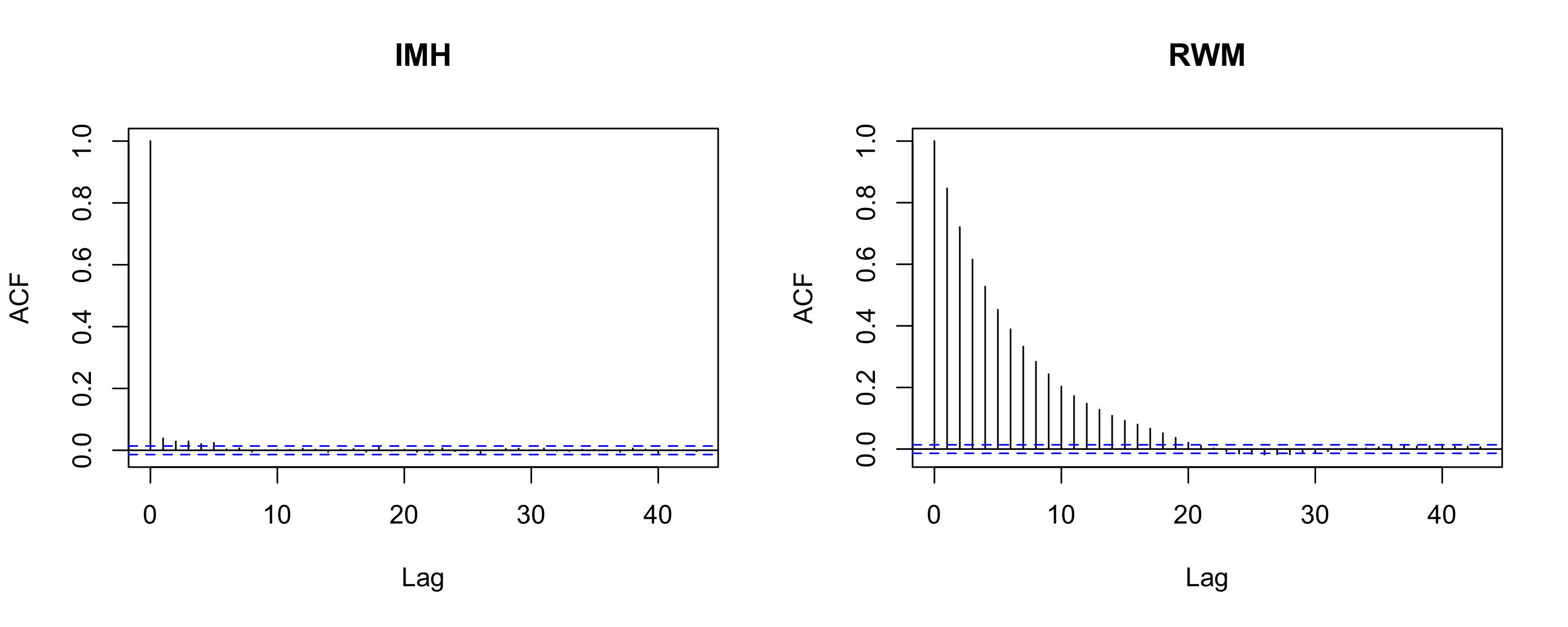}
  \end{subfigure}  
  \caption{Modelling of the warfarin PK data. Top plot: convergence of the empirical medians of $\dens(k_i | y_i ; \theta)$ for a single individual. Comparison between the reference MH algorithm (blue) and the nlme-IMH (red). Bottom plot: Autocorrelation plots of the MCMC samplers for parameter $k_i$.}
  \label{median_acf}
\end{figure}

We run $20\,000$ iterations of these two algorithms and evaluate their convergence by looking at the convergence of the median for the three components of $\psi_i$.
We see Figure~\ref{median_acf} that, for parameter $k_i$, the sequences of empirical median obtained with the two algorithms converge to the same value, which is supposed to be  the theoretical median of the conditional distribution. 
It is interesting to note that the empirical median with the nlme-IMH converge very rapidly. This is interesting in the population approach framework because it is mainly the median values of each conditional distribution that are used to infer the population distribution. 
Autocorrelation plots, Figure~\ref{median_acf}, highlight slower mixing of the RWM whereas samples from the nlme-IMH can be considered independent few iterations after the chain has been initialized.
Comparison for all three dimensions of the individual parameter $\psi_i$ using a Student proposal distribution can be found in Appendix \ref{appendix:numerical.warfa}.

The Mean Square Jump Distance (MSJD) as well as the Effective Sample Size (ESS) of the two methods are reported in Table~\ref{table:msjdandess}.
MSJD is a measure used to diagnose the mixing of the chain. It is calculated as the mean of the squared euclidean distance between every point and its previous point. Usually, this quantity indicates if the chain is moving enough or getting stuck at some region and the ESS is a quantity that estimates the number of independent
samples obtained from a chain. Larger values of those two quantities for our method show greater performance of the sampler in comparison with the RWM.

\begin{table}[h!]
  \begin{center}
    \caption{MSJD and ESS per dimension.}
    \label{table:msjdandess}
    \begin{tabular}{lccccccr} 
      & &  \textbf{MSJD}& & &  & \textbf{ESS} &  \\
            \hline
       & $ka_i$ & $V_i$ & $k_i$& & $ka_i$ & $V_i$ & $k_i$ \\
      \hline
      RWM & 0.009 & 0.002 & 0.006 & & 1728 & 3414 & 3784   \\
      \textbf{nlme-IMH}            & 0.061 & 0.004 & 0.018  & & 13694 & 14907 & 19976 \\
    \end{tabular}
  \end{center}
\end{table}

\textit{Comparison with state-of-the-art methods:}
We then compare our new approach to the three following samplers: an independent sampler that uses variational approximation as proposal distribution \cite{freitas}, the MALA \cite{robertsmala} and the No-U-Turn Sampler \cite{hoffman2014no}. 

The same design and settings (dataset, model parameter estimate, individual) as in section \ref{sec:mcmc_continuous} are used throughout the following experiments.

\subsubsubsection{Variational MCMC algorithm}
The Variational MCMC algorithm \cite{freitas} is a MCMC algorithm with independent proposal. The proposal distribution is a multivariate Gaussian distribution whose parameters are obtained by a variational approach that consists in minimising the Kullback Leibler divergence between a multivariate Gaussian distribution $q(\psi_i,\delta)$, and the target distribution for a given model parameter estimate $\theta$ noted $\dens(\psi_i|y_i,\theta)$. 
This problem boils down to maximizing the so-called Evidence Lower Bound $\ELBO(\theta)$ defined as:
\begin{equation}\label{eq:elbo}
    \ELBO(\delta) \triangleq \int{q(\psi_i,\delta)\left(\log \dens(y_i,\psi_i,\theta) - \log q(\psi_i,\delta)\right)\textrm{d}\psi_i}  \eqs.
\end{equation}
We use the Automatic Differentiation Variational Inference (ADVI) \cite{autom2015} implemented in RStan (R Package \cite{rstan}) to obtain the vector of parameters noted $\delta_{VI}$ defined as:
$$
\delta_{VI} \triangleq \argmax \limits_{\delta \in \mathbb{R}^p \times \mathbb{R}^{p\times p}} \ELBO(\delta) \eqs.
$$
The algorithm stops when the variation of the median of the objective function falls below the $1\%$ threshold. The means and standard deviations of our nlme-IMH and the Variational MCMC proposals compare with the posterior mean (calculated using the NUTS \cite{hoffman2014no}) as follows:

\begin{table}[h!]
  \begin{center}
    \caption{Means and standard deviations.}
    \label{table:means}
    \begin{tabular}{lccccccr} 
      & &  \textbf{Means}& & &  & \textbf{Stds} &  \\
            \hline
       & $ka_i$ & $V_i$ & $k_i$& & $ka_i$ & $V_i$ & $k_i$ \\
      \hline
      Variational proposal & 0.90 & 7.93 & 0.48 & & 0.14 & 0.03 & 0.07  \\
      \textbf{Laplace proposal}           & 0.88 & 7.93 & 0.52 & & 0.18 & 0.04 & 0.09 \\
      NUTS (ground truth)           & 0.91 & 7.93 & 0.51  & & 0.18 &  0.05 &  0.09  \\
    \end{tabular}
  \end{center}
\end{table}
    

We observe that the mean of the variational approximation is slightly shifted from the estimated posterior mode (see table \ref{table:means} for comparison) whereas a considerable difference lies in the numerical value of the covariance matrix obtained with ADVI. The empirical standard deviation of the Variational MCMC proposal is much smaller than our new proposal defined by \eqref{cont:proposalcov} (see table \ref{table:means}), which slows down the MCMC convergence. 

\begin{figure}[thp]
\begin{center}
\includegraphics[width=9cm,keepaspectratio]{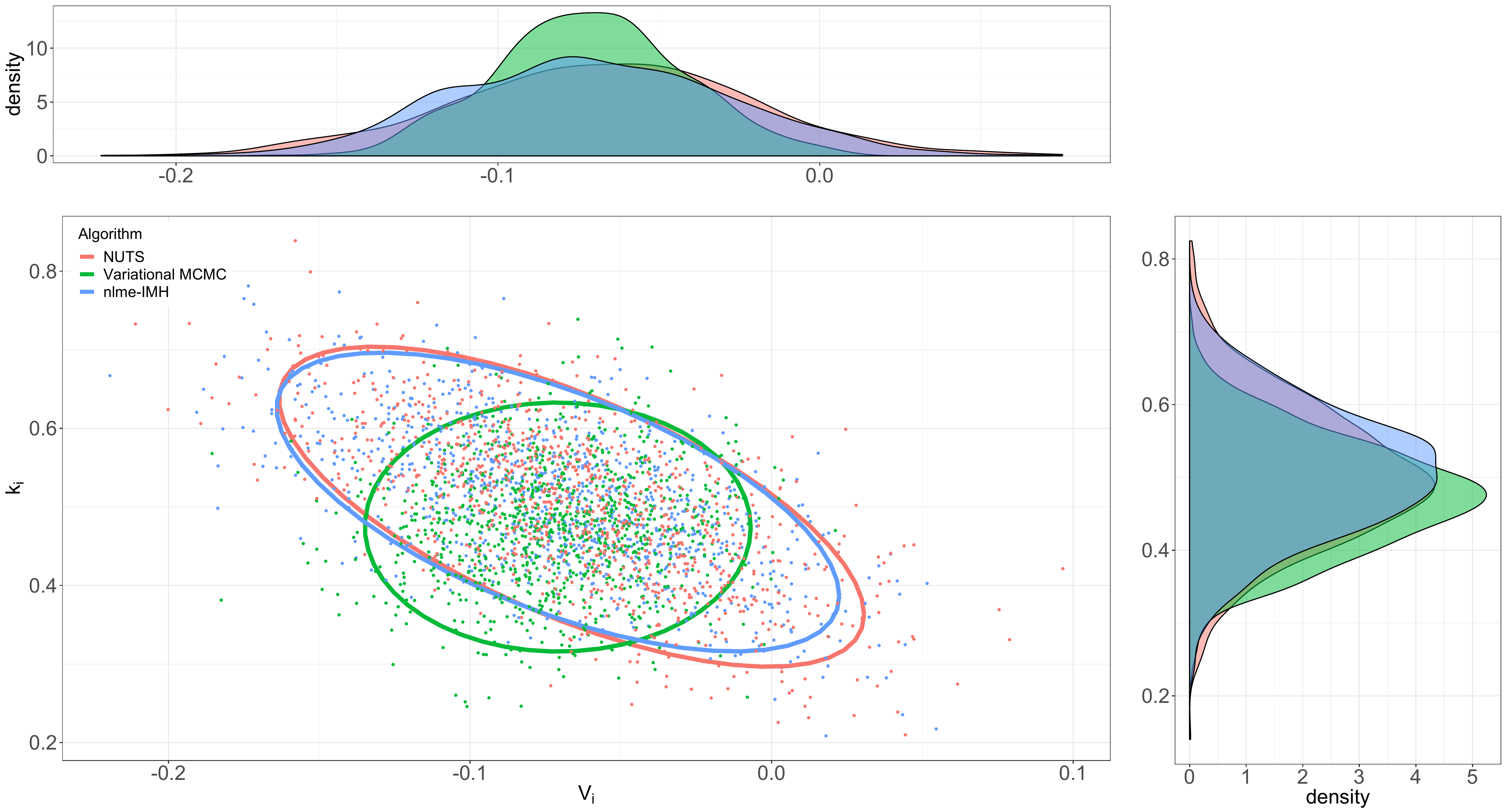}
\caption{Modelling of the warfarin PK data: Comparison between the proposals of the nlme-IMH (blue), the Variational MCMC (green) and the empirical target distribution sampled using the NUTS (red). Marginals and biplots of the conditional distributions $k_i|y_i$ and $V_i|y_i$ for a single individual. Ellipses containing $90 \%$ of the data points are represented on the main plot.}
\label{biplotkV}
\end{center}
\end{figure}

Figure \ref{biplotkV} shows the proposals marginals and the marginal posterior distribution for the individual parameters $k_i$ and $V_i$. Biplot of the samples drawn from the two multivariate Gaussian proposals (our independent proposal and the variational MCMC proposal) as well as samples drawn from the posterior distribution (using the NUTS) are also presented in this figure. We conclude that both marginal and bivariate posterior distributions are better approximated by our independent proposal than the one resulting from a KL divergence optimization.

\begin{figure}[thp]
\begin{center}
\includegraphics[width=\textwidth]{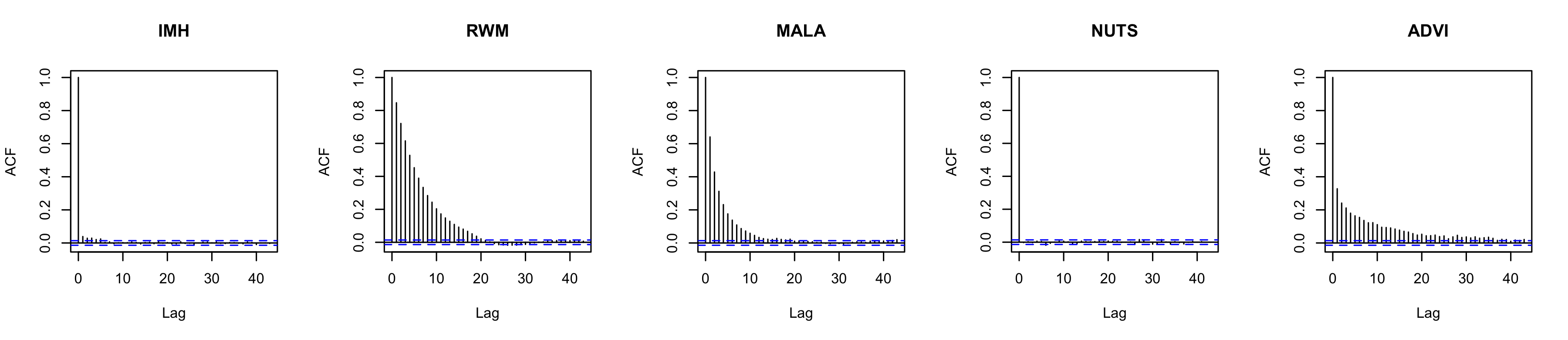}
\caption{Modelling of the warfarin PK data: Autocorrelation plots of the MCMC samplers for parameter $k_i$.}
\label{acf_all}
\end{center}
\end{figure}

Besides similar marginal variances, both our independent proposal and the true posterior share a strong anisotropic nature, confirmed by the similar correlation values of table \ref{table:correlation} (see Appendix~\ref{appendix:numerical.warfa}). Same characteristics are observed for the other parameters.
Those highlighted properties leads to a better performance of the nlme-IMH versus the ADVI sampler as reflected in Figure~\ref{acf_all}. Larger jumps of the chain and bigger ESS show how effective the nlme-IMH is compared to the ADVI (see Table~\ref{table:msjdandess_all}).

\subsubsubsection{Metropolis Adjusted Langevin Algorithm (MALA) and No-U-Turn Sampler (NUTS)}
We now compare our method to the MALA, which proposal is defined by \eqref{eq:update.mala}.
The gradient of the log posterior distribution $\nabla_{\psi_i} \log \dens(\psi_i^{(k)}|y_i)$ is also calculated by Automatic Differentiation.
In this numerical example, the MALA has been initialized at the MAP and the stepsize ($\gamma = 10^{-2}$) is tuned such that the acceptance rate of $0.57$ is reached \cite{roberts}. 

We also compare the implementation of NUTS \cite{hoffman2014no,gelman} in the RStan package to our method in Figure~\ref{acf_all}. 
Figure~\ref{acf_all} highlights good convergence of a well-tuned MALA and the NUTS. nlme-IMH and NUTS mixing properties, from autocorrelation plots in Figure~\ref{acf_all} seem to be similar and much better than all of the other methods.
Table~\ref{table:msjdandess_all} presents a benchmark of those methods regarding MSJD and ESS. 
Both nlme-IMH and NUTS have better performances here. 
For parameters $ka$ and $V$, the ESS of the NUTS, presented as a gold standard sampler for this king of problem, are slightly higher than our proposed method.

\begin{table}[h!]
  \begin{center}
    \caption{MSJD and ESS per dimension.}
    \label{table:msjdandess_all}
    \begin{tabular}{lccccccr} 
      & &  \textbf{MSJD}& & &  & \textbf{ESS} &  \\
            \hline
       & $ka_i$ & $V_i$ & $k_i$& & $ka_i$ & $V_i$ & $k_i$ \\
      \hline
      RWM & 0.009 & 0.002 & 0.006 & & 1728 & 3414 & 3784   \\
      \textbf{nlme-IMH}            & 0.061 & 0.004 & 0.018  & & 13694 & 14907 & 19976 \\
	MALA & 0.024 & 0.002 & 0.006 & & 3458 & 3786 & 3688   \\
      NUTS & 0.063 & 0.004 & 0.018 & & 18684 & 19327 & 19083   \\
      ADVI & 0.037 & 0.002 & 0.010 & & 2499 & 1944 & 2649   \\   
    \end{tabular}
  \end{center}
\end{table}

In practice, those three methods imply tuning phases that are computationally involved , warming up the chain and a careful initialisation whereas our independent sampler is automatic and fast to implement.
Investigating the asymptotic convergence behavior of those methods highlights the competitive properties of our IMH algorithm to sample from the target distribution.

Since our goal is to embed those samplers into a MLE algorithm such as the SAEM, we shall now study how they behave in the very first iterations of the MCMC procedure. Recall that the SAEM requires only few iterations of MCMC sampling under the current model parameter estimate. We present this non asymptotic study in the following section.

\subsubsection{Comparison of the chains for the first 500 iterations}
\begin{figure}[thp]
\begin{center}
\includegraphics[width=\textwidth]{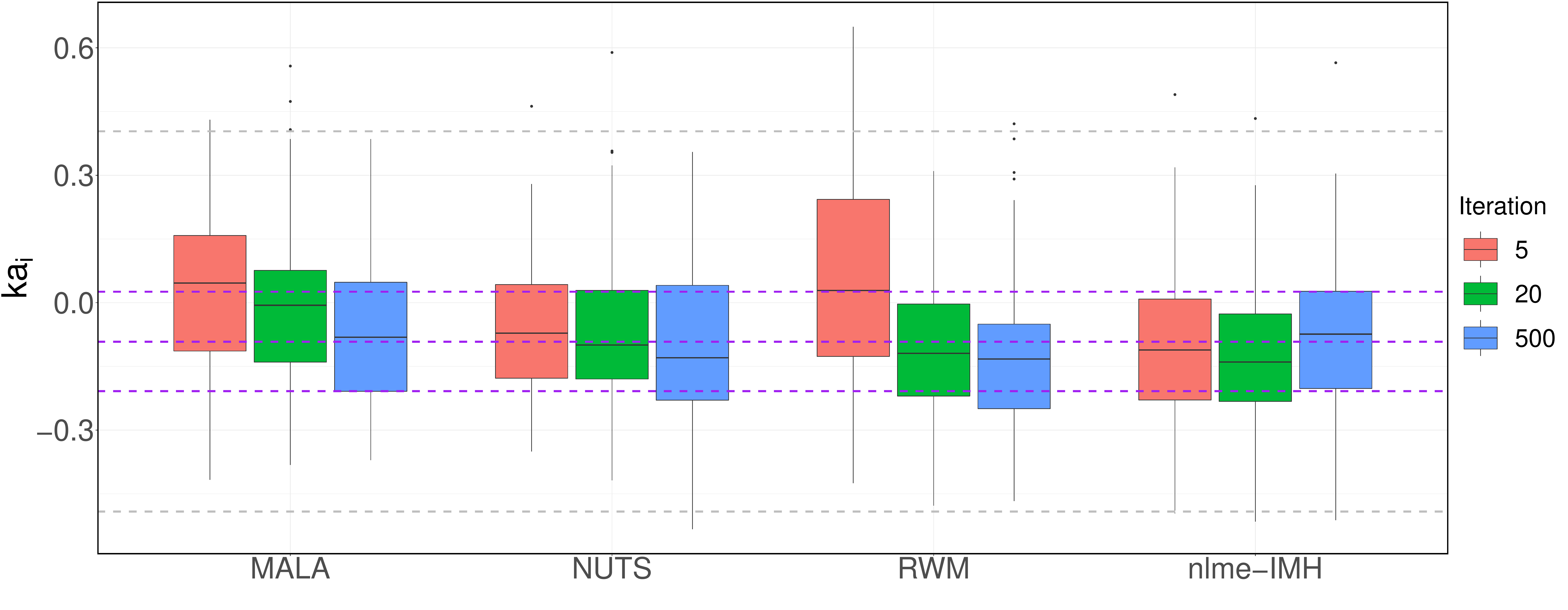}
\caption{Modelling of the warfarin PK data: Boxplots for the RWM, the nlme-IMH, the MALA and the NUTS algorithm, averaged over $100$ independent runs. The groundtruth median, $0.25$ and $0.75$ percentiles are plotted as a dashed purple line and its maximum and minimum as a dashed grey line.}
\label{cont:boxplots}
\end{center}
\end{figure}

We produce $100$ independent runs of the RWM, the nlme-IMH, the MALA and the NUTS for $500$ iterations. The boxplots of the samples drawn at a given iteration threshold (three different thresholds are used) are presented Figure~\ref{cont:boxplots} against the ground truth for the parameter \textbf{ka}. The ground truth has been calculated by running the NUTS for $100\,000$ iterations.

For the three numbers of iteration ($5$,$20$,$500$) considered in Figure~\ref{cont:boxplots}, the median of the nlme-IMH and NUTS samples are closer to the ground truth. Figure~\ref{cont:boxplots} also highlights that all those methods succeed in sampling from the whole distribution after $500$ iterations.

We now use the RWM, the nlme-IMH and the MALA in the SAEM algorithm and observe the convergence of the resulting sequences of parameters.

\subsubsection{Maximum likelihood estimation}

We use the SAEM algorithm to estimate the population PK parameters  $ka_{\rm pop}$, $V_{\rm pop}$ and $k_{\rm pop}$, the standard deviations of the random effects $\omega_{k_a}$, $\omega_{V}$ and $\omega_{k}$ and the residual variance $\sigma^2$.

The stepsize $\gamma_k$ is set to 1 during the first 100 iterations and then decreases as $1/k^{a}$ where $a = 0.7$ during the next 100 iterations. 

Here we compare the standard SAEM algorithm, as implemented in the saemix R package, with the f-SAEM algorithm and the SAEM using the MALA sampler. In this example, the nlme-IMH and the MALA are only used during the first $20$ iterations of the SAEM. The standard MH algorithm is then used.

\begin{figure}[thp]
\begin{center}
\includegraphics[width=\textwidth]{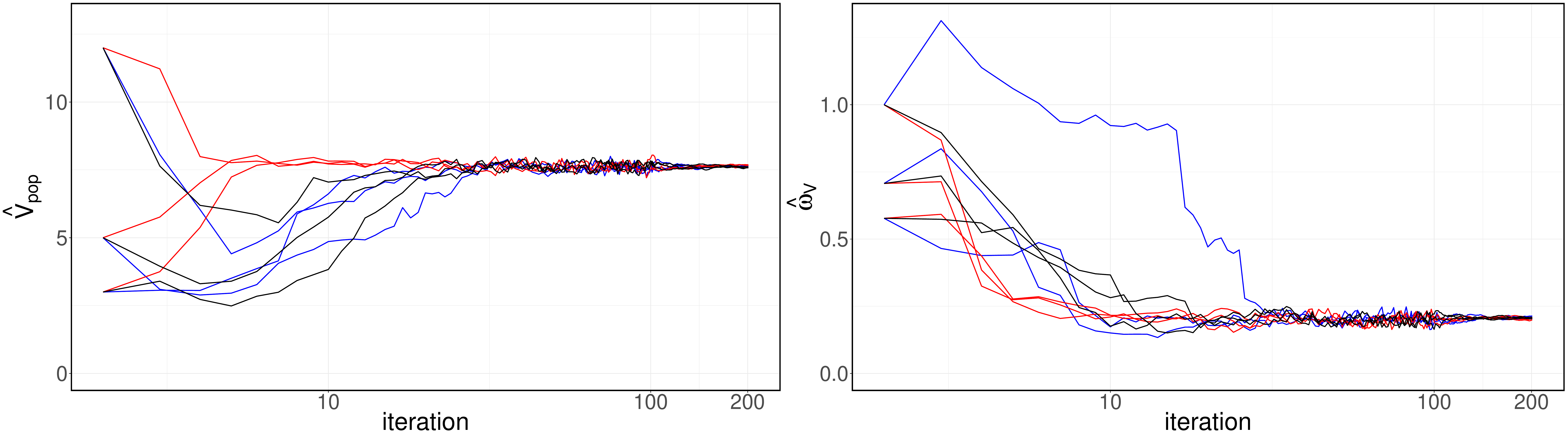}
\caption{Estimation of the population PK parameters for the warfarin data: convergence of the sequences of estimates $(\hat{V}_{{\rm pop},k}, 1\leq k \leq 200)$ and $(\hat{\omega}_{V,k}, 1\leq k \leq 200)$ obtained with SAEM and three different initial values using the reference MH algorithm (blue), the f-SAEM (red) and the SAEM using the MALA sampler (black).} 
\label{realwarfa}
\end{center}
\end{figure}

Figure \ref{realwarfa} shows the estimates of $V_{\rm pop}$ and $\omega_{V}$ computed at each iteration of these three variants of SAEM and starting from three different initial values. First of all, we notice that, whatever the initialisation and the sampling algorithm used, all the runs converge towards the maximum likelihood estimate. It is then very clear that the f-SAEM converges faster than the standard algorithm. 
The SAEM using the MALA algorithm for sampling from the individual conditional distribution presents a similar convergence behavior as the reference. 

We can conclude, for this example, that sampling around the MAP of each individual conditional distribution is the key to a fast convergence of the SAEM during the first iterations. 

\subsubsection{Monte Carlo study}

We conduct a Monte Carlo study to confirm the properties of the f-SAEM algorithm for computing the ML estimates.

$M=50$ datasets have been simulated using the PK model previously used for fitting the warfarin PK data with the following parameter values:
$ka_{\rm pop} =1$, $V_{\rm pop}= 8$, $k_{\rm pop}=0.1$,  $\omega_{ka}=0.5$, $\omega_{V}=0.2$, $\omega_{k}=0.3$ and $\sigma^2=0.5$.
The same original design with $N=32$ patients and a total number of 251 PK measurements were used for all the simulated datasets.
Since all the simulated data are different, the value of the ML estimator varies from one simulation to another. If we run $K$ iterations of SAEM, the last element of the sequence $(\theta_k^{(m)}, 1 \leq k \leq K)$ is the estimate obtained from the $m$-th simulated dataset. 
To investigate how fast $(\theta_k^{(m)}, 1 \leq k \leq K)$ converges to $\theta_K^{(m)}$ we study how fast $(\theta_k^{(m)} - \theta_K^{(m)}, 1 \leq k \leq K)$ goes to 0.
For a given sequence of estimates, we can then define, at each iteration $k$ and for each component $\ell$ of the parameter, the mean square distance over the replicates
\begin{equation}\label{error}
E_k(\ell) = \frac{1}{M}\sum_{m=1}^{M}{\left(\theta_k^{(m)}(\ell) - \theta_K^{(m)}(\ell) \right)^2}\eqs.
\end{equation}

\begin{figure}[thp]
\begin{center}
\includegraphics[width=\textwidth]{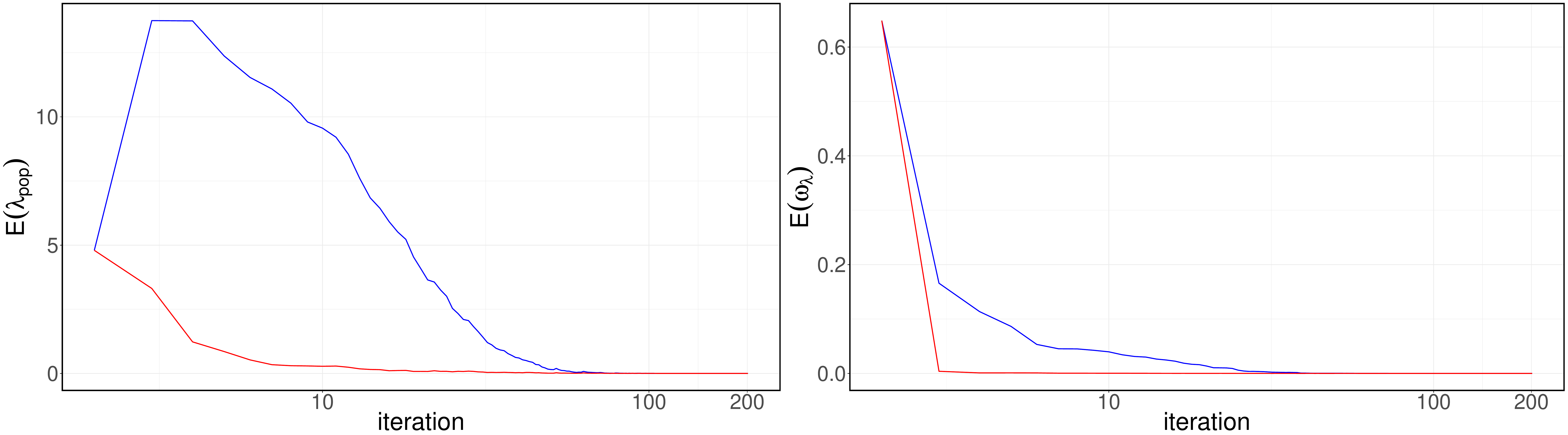}
\caption{Convergence of the sequences of mean square distances $(E_k(V_{\rm pop}), 1\leq k \leq 200)$ and $(E_k(\omega_V), 1\leq k \leq 200)$ for  $V_{\rm pop}$ and $\omega_{V}$ obtained with SAEM on $M=50$ synthetic datasets using the reference MH algorithm (blue) and the f-SAEM (red).}
\label{se_war}
\end{center}
\end{figure}

Figure \ref{se_war} shows using the new proposal leads to a much faster convergence towards the maximum likelihood estimate. Less than 10 iterations are required to converge with the f-SAEM on this example, instead of 50 with the original version. It should also be noted that the distance decreases monotonically. 
The sequence of estimates approaches the target at each iteration, compared to the standard algorithm which makes twists and turns before converging.

\subsection{Time-to-event Data Model}
\subsubsection{The model}
In this section, we consider a Weibull model for time-to-event data \cite{laviellebook, zhang}. For individual $i$, the hazard function of this model is:
\begin{align}\label{weibullmodel}
& h(t, \psi_i) = \frac{\beta_i}{\lambda_i}\left(\frac{t}{\lambda_i}\right)^{\beta_i-1}\eqs.
\end{align}
Here, the vector of individual parameters is $\psi_i = (\lambda_i, \beta_i)$. These two parameters are assumed to be independent and  lognormally distributed:
\begin{align} \label{indivtte}
\log(\lambda_i) \sim \mathcal{N}(\log(\lambda_{\rm pop}), \omega^2_{\lambda}), \log(\beta_i) \sim \mathcal{N}(\log(\beta_{\rm pop}), \omega^2_{\beta})\eqs.
\end{align}
Then, the vector of population parameters is $\theta = (\lambda_{\rm pop}, \beta_{\rm pop}, \omega_{\lambda}, \omega_{\beta})$.

Repeated events were generated, for $N=100$ individuals, using the Weibull model \eqref{weibullmodel} with $\lambda_{\rm pop} = 10$, $\omega_{\lambda} = 0.3$, $\beta_{\rm pop} = 3$ and $\omega_{\beta} = 0.3$ and assuming a right censoring time $\tau_c = 20$.

\subsubsection{MCMC Convergence Diagnostic}
Similarly to the previous section, we start by looking at the behaviour of the MCMC procedure used for sampling from the conditional distribution $\dens(\psi_i | y_i ; \theta)$ for a given individual $i$ and assuming that $\theta$ is known. We use the generating model parameter in these experiments ($\theta = (\lambda_{\rm pop} = 10, \beta_{\rm pop} = 3,\omega_{\lambda} = 0.3 ,\omega_{\beta} = 0.3)$).

\begin{figure}[ht]
  \centering
  \begin{subfigure}{\linewidth}
    \centering
    \includegraphics[width=.8\linewidth]{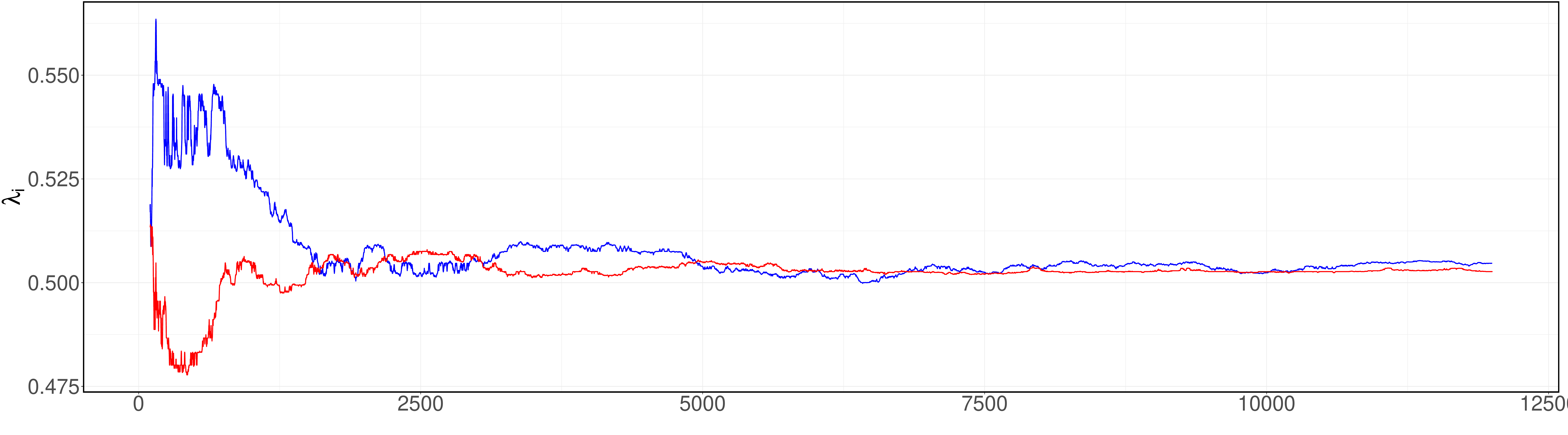}
  \end{subfigure}

  \begin{subfigure}{\linewidth}
    \centering
    \includegraphics[width=.8\linewidth]{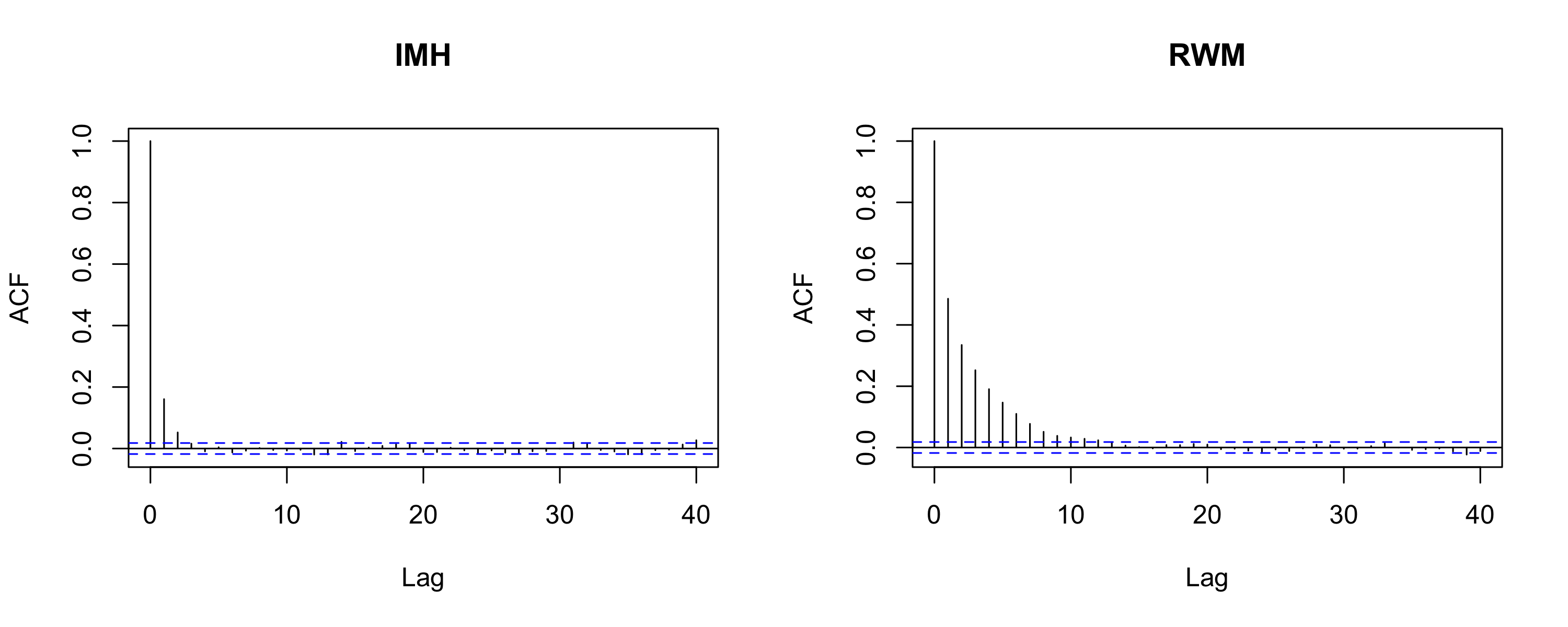}
  \end{subfigure}  
  \caption{Time-to-event data modelling. Top plot: convergence of the empirical medians of $\dens(\lambda_i | y_i ; \theta)$ for a single individual. Comparison between the reference MH algorithm (blue) and the nlme-IMH (red). Bottom plot: Autocorrelation plots of the MCMC samplers for parameter $\lambda_i$.}
  \label{median_acf_rtte}
\end{figure}

We run $12\,000$ iterations of the reference MH algorithm the nlme-IMH to estimate the median of the posterior distribution of $\lambda_i$.
We see Figure~\ref{median_acf_rtte} that the sequences of empirical medians obtained with the two procedures converge to the same value but the new algorithm converges  faster than the standard MH algorithm. Autocorrelation plots, Figure~\ref{median_acf_rtte}, are also significantly showing the advantage of the new sampler as the chain obtained with the nlme-IMH is mixing almost ten times faster than the reference sampler.

\begin{table}[h!]
  \begin{center}
    \caption{MSJD and ESS per dimension.}
    \label{table:msjdandess}
    \begin{tabular}{lccccr} 
      &  \textbf{MSJD} & & &  \textbf{ESS} &    \\
            \hline
       & $\lambda_i$ & $\beta_i$ & & $\lambda_i$ & $\beta_i$  \\
      \hline
      RWM & 0.055 & 0.093  & &  3061 & 1115    \\
      \textbf{nlme-IMH}            & 0.095 & 0.467   & & 8759 & 8417  \\
    \end{tabular}
  \end{center}
\end{table}
Plots for the other parameter can be found in Appendix \ref{appendix:numerical.tte}.
Comparisons with state-of-the-art methods were conducted as in the previous section.
These comparisons led us to the same remarks as those made for the previous continuous data model both on the asymptotic and non asymptotic regimes.

\subsubsection{Maximum likelihood estimation of the population parameters}

We run the standard SAEM algorithm implemented in the saemix package
(extension of this package for noncontinuous data models is available on GitHub:  \textsf{https://github.com/belhal/saemix}) and the f-SAEM on the generated dataset.

\begin{figure}[thp]
\begin{center}
\includegraphics[width=\textwidth]{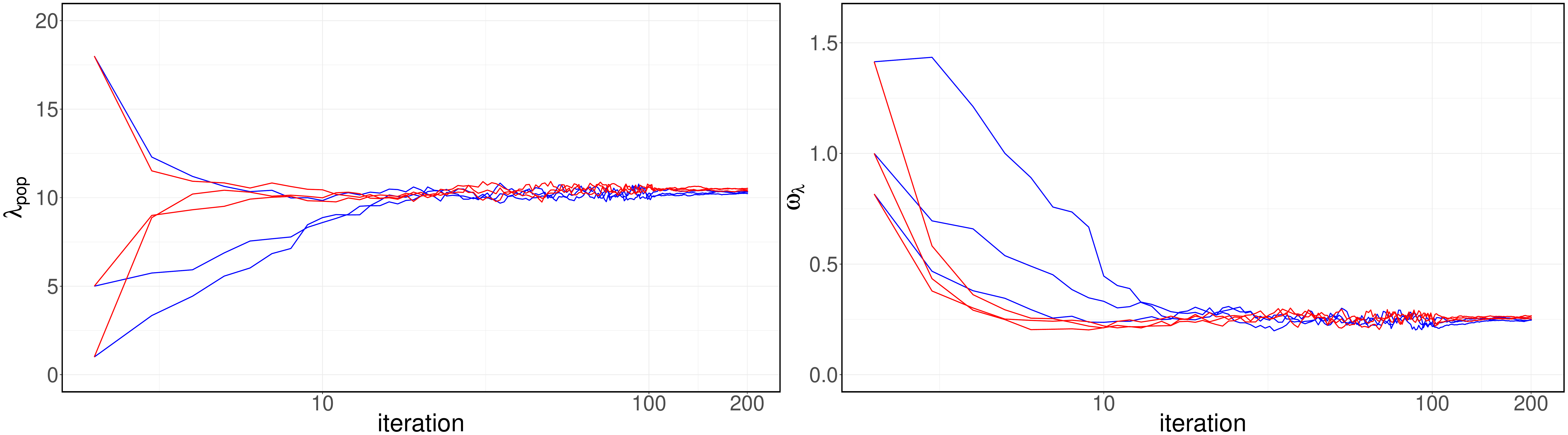}
\caption{Population parameter estimation in time-to-event-data models: convergence of the sequences of estimates $(\hat{\lambda}_{{\rm pop},k}, 1\leq k \leq 200)$ and $(\hat{\omega}_{\lambda,k}, 1\leq k \leq 200)$ obtained with SAEM and three different initial values using the reference MH algorithm (blue) and the f-SAEM (red).}
\label{rtte_init}
\end{center}
\end{figure}

Figure~\ref{rtte_init} shows the estimates of $\lambda_{ \rm pop}$ and $\omega_{\lambda}$ computed at each iteration of the two versions of the SAEM and starting from three different initial values. 
The same behaviour is observed as in the continuous case: regardless the initial values and the algorithm, all the runs converge to the same solution but convergence is much faster with the proposed method. The same comment applies for the two other parameters $\beta_{ \rm pop}$ and $\omega_{\beta}$.

\subsubsection{Monte Carlo study}
\begin{figure}[thp]
\begin{center}
\includegraphics[width=\textwidth]{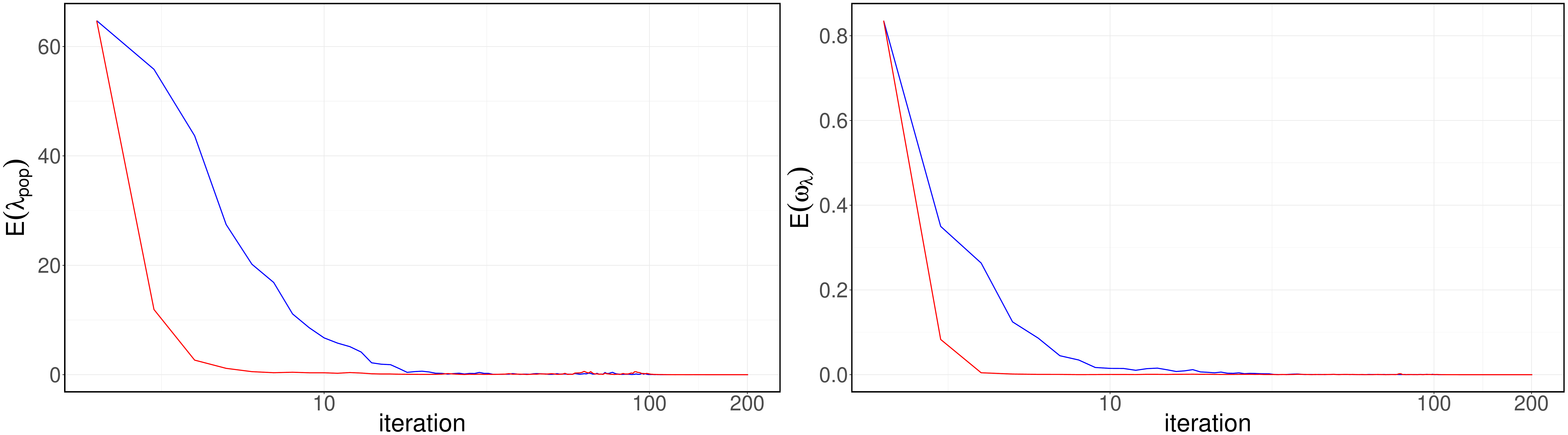}
\caption{Convergence of the sequences of mean square distances $(E_k(\lambda_{\rm pop}), 1\leq k \leq 200)$ and $(E_k(\omega_{\lambda}), 1\leq k \leq 200)$ for  $\lambda_{\rm pop}$ and $\omega_{\lambda}$ obtained with SAEM from $M=50$ synthetic datasets using the reference MH algorithm (blue) and the f-SAEM (red).}
\label{se_rtte}
\end{center}
\end{figure}

We now conduct a Monte Carlo study in order to confirm the good properties of the new version of the SAEM algorithm for estimating the population parameters of a time-to-event data model. 
$M=50$ synthetic datasets are generated using the same design as above. Figure~\ref{se_rtte} shows the convergence of the mean square distances defined in \eqref{error} for $\lambda_{\rm pop}$ and $\omega_{\lambda}$. All these distances converge monotonically to 0 which means that both algorithms properly converge to the maximum likelihood estimate, but very few iterations are required  with the new version to converge while about thirty iterations are needed with the SAEM.

\section{Conclusion and discussion}
We present in this article an independent Metropolis-Hastings procedure for sampling random effects from their conditional distributions and a fast MLE algorithm, called the f-SAEM, in nonlinear mixed effects models.
 
The idea of the method is to approximate each individual conditional distribution by a multivariate normal distribution. A Laplace approximation makes it possible to consider any type of data, but we have shown that, in the case of continuous data, this approximation is equivalent to linearizing the structural model around the conditional mode of the random effects.

The numerical experiments demonstrate that the proposed nlme-IMH sampler converges faster to the target distribution than a standard random walk Metropolis. This practical behaviour is partly explained by the fact that the conditional mode of the random effects in the linearized model coincides with the conditional mode of the random effects in the original model. 
The proposal distribution is therefore a normal distribution centered around this MAP.
On the other hand, the dependency structure in the conditional distribution of the random effects is well approximated by the covariance structure of the Gaussian proposal.
So far, we have mainly applied our method to standard problems encountered in pharmacometrics and for which the number of random effects remains small. 
It can nevertheless be interesting to see how this method behaves in higher dimension and compare it with methods adapted to such situations such as MALA or HMC.
Lastly, we have shown that this new IMH algorithm can easily be embedded in the SAEM algorithm for maximum likelihood estimation of the population parameters.  
Our numerical studies have shown empirically that the new transition kernel is effective in the very first iterations. It is of great interest to determine automatically and in an adaptive way an optimal scheme of kernel transitions combining this new proposal with the block-wise random walk Metropolis.

\newpage
\bibliographystyle{plainnat}
\bibliography{ref}

\begin{thebibliography}{49}
\providecommand{\natexlab}[1]{#1}
\providecommand{\url}[1]{\texttt{#1}}
\expandafter\ifx\csname urlstyle\endcsname\relax
  \providecommand{\doi}[1]{doi: #1}\else
  \providecommand{\doi}{doi: \begingroup \urlstyle{rm}\Url}\fi

\bibitem[Agresti(1990)]{agresti}
Alan Agresti.
\newblock \emph{Categorical data analysis}.
\newblock A Wiley-Interscience publication. Wiley, New York, 1990.

\bibitem[Allassonniere and Kuhn(2013)]{kuhnamala}
St{\'e}phanie Allassonniere and Estelle Kuhn.
\newblock {Convergent Stochastic Expectation Maximization algorithm with
  efficient sampling in high dimension. Application to deformable template
  model estimation}.
\newblock \emph{arXiv preprint arXiv:1207.5938}, 2013.

\bibitem[Andersen(2006)]{andersen}
P.~K. Andersen.
\newblock {Survival Analysis}.
\newblock \emph{Wiley Reference Series in Biostatistics}, 2006.

\bibitem[Andrieu and Thoms(2008)]{andrieu2008tutorial}
Christophe Andrieu and Johannes Thoms.
\newblock A tutorial on adaptive mcmc.
\newblock \emph{Statistics and computing}, 18\penalty0 (4):\penalty0 343--373,
  2008.

\bibitem[Andrieu et~al.(2006)Andrieu, Moulines, et~al.]{andrieu2006ergodicity}
Christophe Andrieu, {\'E}ric Moulines, et~al.
\newblock On the ergodicity properties of some adaptive mcmc algorithms.
\newblock \emph{The Annals of Applied Probability}, 16\penalty0 (3):\penalty0
  1462--1505, 2006.

\bibitem[Andrieu et~al.(2009)Andrieu, Roberts, et~al.]{andrieu2009pseudo}
Christophe Andrieu, Gareth~O Roberts, et~al.
\newblock The pseudo-marginal approach for efficient monte carlo computations.
\newblock \emph{The Annals of Statistics}, 37\penalty0 (2):\penalty0 697--725,
  2009.

\bibitem[Andrieu et~al.(2010)Andrieu, Doucet, and
  Holenstein]{andrieu2010particle}
Christophe Andrieu, Arnaud Doucet, and Roman Holenstein.
\newblock Particle markov chain monte carlo methods.
\newblock \emph{Journal of the Royal Statistical Society: Series B (Statistical
  Methodology)}, 72\penalty0 (3):\penalty0 269--342, 2010.

\bibitem[Atchadé and Rosenthal(2005)]{atchade}
Yves~F. Atchadé and Jeffrey~S. Rosenthal.
\newblock {On adaptive Markov chain Monte Carlo algorithms}.
\newblock \emph{Bernoulli}, 11\penalty0 (5):\penalty0 815--828, 10 2005.
\newblock \doi{10.3150/bj/1130077595}.

\bibitem[Beal and Sheiner(1980)]{beal}
Stuart Beal and Lewis Sheiner.
\newblock The {NONMEM} system.
\newblock \emph{The American Statistician}, 34\penalty0 (2):\penalty0 118--119,
  1980.

\bibitem[Betancourt(2017)]{betancourt2017conceptual}
Michael Betancourt.
\newblock {A conceptual introduction to Hamiltonian Monte Carlo}.
\newblock \emph{arXiv preprint arXiv:1701.02434}, 2017.

\bibitem[Brooks et~al.(2011)Brooks, Gelman, Jones, and
  Meng]{brooks2011handbook}
Steve Brooks, Andrew Gelman, Galin Jones, and Xiao-Li Meng.
\newblock \emph{Handbook of markov chain monte carlo}.
\newblock CRC press, 2011.

\bibitem[Brosse et~al.(2017)Brosse, Durmus, Moulines, and
  Sabanis]{brosse2017tamed}
Nicolas Brosse, Alain Durmus, {\'E}ric Moulines, and Sotirios Sabanis.
\newblock The tamed unadjusted langevin algorithm.
\newblock \emph{arXiv preprint arXiv:1710.05559}, 2017.

\bibitem[Carpenter et~al.(2017)Carpenter, Gelman, Hoffman, Lee, Ben,
  Betancourt, Brubaker, Guo, Li, and Riddell]{gelman}
Bob Carpenter, Andrew Gelman, Matthew Hoffman, Daniel Lee, Goodrich Ben,
  Michael Betancourt, Marcus Brubaker, Jiqiang Guo, Peter Li, and Allen
  Riddell.
\newblock Stan: {A} probabilistic programming language.
\newblock \emph{Journal of Statistical Software}, 76\penalty0 (1), 2017.

\bibitem[Chan et~al.(2011)Chan, Jacqmin, Lavielle, McFadyen, and
  Weatherley]{chan}
P.~L.~S. Chan, P.~Jacqmin, M.~Lavielle, L.~McFadyen, and B.~Weatherley.
\newblock The use of the {SAEM} algorithm in {MONOLIX} software for estimation
  of population pharmacokinetic-pharmacodynamic-viral dynamics parameters of
  maraviroc in asymptomatic {HIV} subjects.
\newblock \emph{Journal of Pharmacokinetics and Pharmacodynamics}, 38\penalty0
  (1):\penalty0 41--61, 2011.

\bibitem[Comets et~al.(2017)Comets, Lavenu, and Lavielle]{comets}
Emmanuelle Comets, Audrey Lavenu, and Marc Lavielle.
\newblock Parameter estimation in nonlinear mixed effect models using saemix,
  an r implementation of the saem algorithm.
\newblock \emph{Journal of Statistical Software}, 80\penalty0 (3):\penalty0
  1--42, 2017.

\bibitem[de~Freitas et~al.(2001)de~Freitas, H{\o}jen-S{\o}rensen, Jordan, and
  Russell]{freitas}
Nando de~Freitas, Pedro H{\o}jen-S{\o}rensen, Michael~I Jordan, and Stuart
  Russell.
\newblock Variational mcmc.
\newblock \emph{Proceedings of the Seventeenth Conference on Uncertainty in
  Artificial Intelligence}, pages 120--127, 2001.

\bibitem[Delyon et~al.(1999)Delyon, Lavielle, and Moulines]{lavielle}
Bernard Delyon, Marc Lavielle, and Eric Moulines.
\newblock {Convergence of a stochastic approximation version of the EM
  algorithm}.
\newblock \emph{Ann. Statist.}, 27\penalty0 (1):\penalty0 94--128, 03 1999.
\newblock \doi{10.1214/aos/1018031103}.

\bibitem[Donnet and Samson(2013)]{donnet2013using}
Sophie Donnet and Adeline Samson.
\newblock Using pmcmc in em algorithm for stochastic mixed models: theoretical
  and practical issues.
\newblock \emph{Journal de la Soci{\'e}t{\'e} Fran{\c{c}}aise de Statistique},
  155\penalty0 (1):\penalty0 49--72, 2013.

\bibitem[Doucet et~al.(2000)Doucet, Godsill, and Andrieu]{doucet2000sequential}
Arnaud Doucet, Simon Godsill, and Christophe Andrieu.
\newblock On sequential monte carlo sampling methods for bayesian filtering.
\newblock \emph{Statistics and computing}, 10\penalty0 (3):\penalty0 197--208,
  2000.

\bibitem[Griewank and Walther(2008)]{griewank2008evaluating}
Andreas Griewank and Andrea Walther.
\newblock \emph{Evaluating derivatives: principles and techniques of
  algorithmic differentiation}, volume 105.
\newblock Siam, 2008.

\bibitem[Haario et~al.(2001)Haario, Saksman, Tamminen,
  et~al.]{haario2001adaptive}
Heikki Haario, Eero Saksman, Johanna Tamminen, et~al.
\newblock An adaptive metropolis algorithm.
\newblock \emph{Bernoulli}, 7\penalty0 (2):\penalty0 223--242, 2001.

\bibitem[Hoffman and Gelman(2014)]{hoffman2014no}
Matthew~D Hoffman and Andrew Gelman.
\newblock The {No-U-turn} sampler: adaptively setting path lengths in
  {Hamiltonian Monte Carlo}.
\newblock \emph{Journal of Machine Learning Research}, 15\penalty0
  (1):\penalty0 1593--1623, 2014.

\bibitem[Kucukelbir et~al.(2015)Kucukelbir, Ranganath, Gelman, and
  Blei]{autom2015}
Alp Kucukelbir, Rajesh Ranganath, Andrew Gelman, and David Blei.
\newblock Automatic variational inference in stan.
\newblock In C.~Cortes, N.~D. Lawrence, D.~D. Lee, M.~Sugiyama, and R.~Garnett,
  editors, \emph{Advances in Neural Information Processing Systems 28}, pages
  568--576. Curran Associates, Inc., 2015.

\bibitem[Kuhn and Lavielle(2004)]{kuhn}
Estelle Kuhn and Marc Lavielle.
\newblock {Coupling a stochastic approximation version of EM with an MCMC
  procedure}.
\newblock \emph{ESAIM: Probability and Statistics}, 8:\penalty0 115--131, 2004.

\bibitem[Lavielle(2014)]{laviellebook}
Marc Lavielle.
\newblock \emph{Mixed effects models for the population approach: models,
  tasks, methods and tools}.
\newblock CRC press, 2014.

\bibitem[Lavielle and Ribba(2016)]{lavielle2016enhanced}
Marc Lavielle and Benjamin Ribba.
\newblock Enhanced method for diagnosing pharmacometric models: random sampling
  from conditional distributions.
\newblock \emph{Pharmaceutical research}, 33\penalty0 (12):\penalty0
  2979--2988, 2016.

\bibitem[Louis(1982)]{louis}
Thomas~A. Louis.
\newblock Finding the observed information matrix when using the em algorithm.
\newblock \emph{Journal of the Royal Statistical Society, Series B:
  Methodological}, 44:\penalty0 226--233, 1982.

\bibitem[Mbogning et~al.(2015)Mbogning, Bleakley, and Lavielle]{mbogning}
Cyprien Mbogning, Kevin Bleakley, and Marc Lavielle.
\newblock {Joint modeling of longitudinal and repeated time-to-event data using
  nonlinear mixed-effects models and the SAEM algorithm}.
\newblock \emph{{Journal of Statistical Computation and Simulation}},
  85\penalty0 (8):\penalty0 1512--1528, 2015.
\newblock \doi{10.1080/00949655.2013.878938}.

\bibitem[McLachlan and Krishnan(2007)]{mclachlan2007algorithm}
Geoffrey McLachlan and Thriyambakam Krishnan.
\newblock \emph{The EM algorithm and extensions}, volume 382.
\newblock John Wiley \& Sons, 2007.

\bibitem[Mengersen and Tweedie(1996)]{mengersen}
K.~L. Mengersen and R.~L. Tweedie.
\newblock Rates of convergence of the hastings and metropolis algorithms.
\newblock \emph{Ann. Statist.}, 24\penalty0 (1):\penalty0 101--121, 02 1996.
\newblock \doi{10.1214/aos/1033066201}.

\bibitem[Metropolis et~al.(1953)Metropolis, Rosenbluth, Rosenbluth, Teller, and
  Teller]{metropolis}
Nicholas Metropolis, Arianna~W. Rosenbluth, Marshall~N. Rosenbluth, Augusta~H.
  Teller, and Edward Teller.
\newblock Equation of state calculations by fast computing machines.
\newblock \emph{The Journal of Chemical Physics}, 21\penalty0 (6):\penalty0
  1087--1092, 1953.
\newblock \doi{10.1063/1.1699114}.

\bibitem[Migon et~al.(2014)Migon, Gamerman, and Louzada]{gamerman}
H.S. Migon, D.~Gamerman, and F.~Louzada.
\newblock \emph{Statistical Inference: An Integrated Approach, Second Edition}.
\newblock Chapman \& Hall/CRC Texts in Statistical Science. Taylor \& Francis,
  2014.
\newblock ISBN 9781439878804.

\bibitem[Neal et~al.(2011)]{neal2011mcmc}
Radford~M Neal et~al.
\newblock Mcmc using hamiltonian dynamics.
\newblock \emph{Handbook of Markov Chain Monte Carlo}, 2\penalty0
  (11):\penalty0 2, 2011.

\bibitem[O'Reilly and Aggeler(1968)]{oreilly}
Robert~A. O'Reilly and Paul~M. Aggeler.
\newblock Studies on coumarin anticoagulant drugs initiation of warfarin
  therapy without a loading dose.
\newblock \emph{Circulation}, 38\penalty0 (1):\penalty0 169--177, 1968.

\bibitem[Pav(2016)]{pav2016madness}
Steven~E Pav.
\newblock Madness: a package for multivariate automatic differentiation.
\newblock 2016.

\bibitem[Robert and Casella(2010)]{mh:robert}
Christian~P. Robert and George Casella.
\newblock \emph{Metropolis--Hastings Algorithms}, pages 167--197.
\newblock Springer New York, New York, NY, 2010.

\bibitem[Roberts and Rosenthal(1997)]{roberts}
G.~O. Roberts and J.~S. Rosenthal.
\newblock Optimal scaling of discrete approximations to langevin diffusions.
\newblock \emph{J. R. Statist. Soc. B}, 60:\penalty0 255--268, 1997.

\bibitem[Roberts et~al.(1997)Roberts, Gelman, and Gilks]{robertsoptimal}
G.~O. Roberts, A.~Gelman, and W.~R. Gilks.
\newblock Weak convergence and optimal scaling of random walk metropolis
  algorithms.
\newblock \emph{Ann. Appl. Probab.}, 7\penalty0 (1):\penalty0 110--120, 02
  1997.
\newblock \doi{10.1214/aoap/1034625254}.

\bibitem[Roberts and Rosenthal(2011)]{roberts2011quantitative}
Gareth~O Roberts and Jeffrey~S Rosenthal.
\newblock Quantitative non-geometric convergence bounds for independence
  samplers.
\newblock \emph{Methodology and Computing in Applied Probability}, 13\penalty0
  (2):\penalty0 391--403, 2011.

\bibitem[Roberts and Tweedie(1996)]{robertsmala}
Gareth~O. Roberts and Richard~L. Tweedie.
\newblock Exponential convergence of langevin distributions and their discrete
  approximations.
\newblock \emph{Bernoulli}, 2\penalty0 (4):\penalty0 341--363, 12 1996.

\bibitem[Rue et~al.(2009)Rue, Martino, and Chopin]{rue2009approximate}
H{\aa}vard Rue, Sara Martino, and Nicolas Chopin.
\newblock Approximate bayesian inference for latent gaussian models by using
  integrated nested laplace approximations.
\newblock \emph{Journal of the royal statistical society: Series b (statistical
  methodology)}, 71\penalty0 (2):\penalty0 319--392, 2009.

\bibitem[Savic et~al.(2011)Savic, Mentr\'{e}, and Lavielle]{savic}
R.~M. Savic, F.~Mentr\'{e}, and M.~Lavielle.
\newblock Implementation and evaluation of the {SAEM} algorithm for
  longitudinal ordered categorical data with an illustration in
  pharmacokinetics-pharmacodynamics.
\newblock \emph{The AAPS Journal}, 13\penalty0 (1):\penalty0 44--53, 2011.

\bibitem[{Stan Development Team}(2018)]{rstan}
{Stan Development Team}.
\newblock {RStan}: the {R} interface to {Stan}, 2018.
\newblock URL \url{http://mc-stan.org/}.
\newblock R package version 2.17.3.

\bibitem[Stramer and Tweedie(1999)]{stramer}
O.~Stramer and R.~L. Tweedie.
\newblock Langevin-type models i: Diffusions with given stationary
  distributions and their discretizations*.
\newblock \emph{Methodology And Computing In Applied Probability}, 1\penalty0
  (3):\penalty0 283--306, Oct 1999.
\newblock ISSN 1573-7713.
\newblock \doi{10.1023/A:1010086427957}.

\bibitem[Titsias and Papaspiliopoulos(2018)]{titsias2018agrad}
Michalis~K. Titsias and Omiros Papaspiliopoulos.
\newblock Auxiliary gradient‐based sampling algorithms.
\newblock \emph{Journal of the Royal Statistical Society: Series B (Statistical
  Methodology)}, 0\penalty0 (0), 2018.
\newblock \doi{10.1111/rssb.12269}.

\bibitem[Verbeke(1997)]{verbeke1997linear}
Geert Verbeke.
\newblock \emph{Linear mixed models for longitudinal data}.
\newblock Springer, 1997.

\bibitem[Vihola(2012)]{vihola2012robust}
Matti Vihola.
\newblock Robust adaptive metropolis algorithm with coerced acceptance rate.
\newblock \emph{Statistics and Computing}, 22\penalty0 (5):\penalty0 997--1008,
  2012.

\bibitem[Wang(2007)]{wang}
Yaning Wang.
\newblock Derivation of various nonmem estimation methods.
\newblock \emph{Journal of Pharmacokinetics and pharmacodynamics}, 34\penalty0
  (5):\penalty0 575--593, 2007.

\bibitem[Zhang(2016)]{zhang}
Z.~Zhang.
\newblock {Parametric regression model for survival data: Weibull regression
  model as an example}.
\newblock \emph{Ann Transl Med.}, 24, 2016.

\end{thebibliography}

\begin{appendices}
\section{Extensions of model \eqref{indiv_param1}}\label{appendix:modelextensions}
Several extensions of model \eqref{indiv_param1} are also possible. We can assume for instance that the transformed individual parameters are normally distributed:
\begin{equation} \label{indiv_param2}
u(\psi_i) = u(\psipop)+\eta_i\eqs,
\end{equation}
where $u$ is a strictly monotonic transformation applied on the individual parameters $\psi_i$. Examples of such transformation are the logarithmic function (in which case the components of $\psi_i$ are log-normally distributed), the logit and the probit transformations \cite{laviellebook}.  In the following, we either use the original parameter $\psi_i$ or the  Gaussian transformed parameter $u(\psi_i)$.

Another extension of model \eqref{indiv_param1} consists in introducing individual covariates in order to explain part of the inter-individual variability:
\begin{equation}
u(\psi_i) = u(\psipop) + C_i \beta+ \eta_i\eqs,
\end{equation}
where $C_i$ is a matrix of individual covariates. Here, the fixed effects are the vector of coefficients $\beta$ and the vector of typical parameters $\psipop$.

\section{Calculus of the proposal in the noncontinuous case} \label{appendix:noncontinuous}

Laplace approximation (see \cite{gamerman}) consists in approximating an integral of the form
\begin{equation}
I := \int{\exponential^{v(x)}\textrm{d}x}\eqs,
\end{equation}
where $v$ is at least twice differentiable.

The following second order Taylor expansion of the function $v$ around a point $x_0$ 
\begin{equation}
v(x) \approx v(x_0) + \nabla v(x_0)(x-x_0) +\frac{1}{2}(x-x0)\nabla^2v(x_0)(x-x0)\eqs,
\end{equation}
provides an approximation of the integral $I$ (consider a multivariate Gaussian probability distribution function which integral sums to 1):
\begin{equation}
I \approx \exponential^{v(x_0)}\sqrt{\frac{(2\pi)^p}{|-\nabla^2v(x_0)|}} 
\exp\left\{-\frac{1}{2}\trans{\nabla v(x_0)}\nabla^2v(x_0)^{-1}\nabla v(x_0) \right\}\eqs.
\end{equation}

In our context, we can  write the marginal pdf $\dens(y_i)$ that we aim to approximate as
\begin{align}
\dens(y_i) &= \int{\dens(y_i,\psi_i)\textrm{d}\psi_i}\\
& = \int{\exponential^{\log \dens(y_i,\psi_i)}\textrm{d}\psi_i}\eqs.
\end{align}

Then, let
\begin{align}
v(\psi_i) &:= 
\log \dens(y_i,\psi_i)  \\
& = l(\psi_i) + \log \dens(\psi_i)\eqs,
\end{align}
and compute its Taylor expansion around the MAP $\map$. 
We have by definition that
$$\nabla \log \dens(y_i,\map)=0 \eqs,$$
which leads to the following Laplace approximation of $\log \dens(y_i)$:
$$
-2\log \dens(y_i)  \approx -p\log2\pi - 2\log \dens(y_i,\map) + \log \left(\left|-\nabla^2 \log \dens(y_i,\map)\right|\right)\eqs.
$$

We thus obtain the following approximation of the logarithm of the conditional pdf of $\psi_i$ given $y_i$ evaluated at $\map$:
 $$
\log \dens(\map|y_i) \approx -\frac{p}{2}\log2\pi  -\frac{1}{2}\log \left(\left|-\nabla^2 \log \dens(y_i,\map) \right|\right)\eqs,
$$
which is precisely the log-pdf of a multivariate Gaussian distribution with mean $\map$ and  variance-covariance $-\nabla^2 \log \dens(y_i,\map)^{-1}$ with:

\begin{align}
\nabla^2 \log \dens(y_i,\map) &= \nabla^2 \log \dens(y_i|\map) + \nabla^2 \log \dens(\map) \\
&= \nabla^2  l(\map) + \Omega^{-1}\eqs.
\end{align}

\section{Linear continuous data models} \label{appendix:linearmodel}

Let $y_i = \trans{(y_{i,1}, \ldots, y_{i,n_i})}$ and 
$\varepsilon_i =  \trans{(\varepsilon_{i,1}, \ldots, \varepsilon_{i,n_i})}$.
Assume a linear relationship between the observations $y_i$ and the vector of individual parameters $\psi_i$:
\begin{equation} \label{y_lmem}
y_{i} = A_i\psi_i + \varepsilon_{i}\eqs,
\end{equation}
where $A_i \in \mathbb{R}^{n_i \times p}$ is the design matrix for individual $i$, $\psi_i$ is normally distributed with mean $m_i \in \mathbb{R}^{p}$ and covariance $\Omega \in \mathbb{R}^{p \times p}$. Then, the conditional distribution of $\psi_i$ given $y_i$ is a normal distribution with mean $\mu_i$ and variance-covariance matrix $\Gamma_i$ defined as:
\begin{equation} \label{map_lin}
\mu_i = \Gamma_i \left( \frac{\trans{A}_i y_i}{\sigma^2} + \Omega^{-1}m_i\right) \quad \textrm{where} \quad \Gamma_i = \left( \frac{\trans{A}_i A_i}{\sigma^2} + \Omega^{-1}\right)^{-1} \\
\end{equation}
Here, $\mu_i$ is the mode of the conditional distribution of $\psi_i$, known as the Maximum A Posteriori (MAP) estimate, or the Empirical Bayes Estimate (EBE) of $\psi_i$.

\section{Conditional mode under the linearised model}\label{appendix:conditionalmode}
Using \eqref{mapdef2}, $\map$ satisfies:
\begin{equation}\label{eq:conditionMAP}
-\frac{\trans{\jacob_{f_i(\map)}} }{\sigma^2}\left(y_i - f_i(\map)\right) + \Omega^{-1}(\map-m_i) = 0\eqs,
\end{equation}
which leads to the definition of the conditional mean $\mu_i$ of $\psi_i$ given $z_i$, under the linearized model, by:
\begin{align}\label{cont:proposalmean}
\mu_i & = \Gamma_i \frac{\trans{\jacob_{f_i(\map)}} }{\sigma^2} \left(y_i - f_i(\map) + \jacob_{f_i(\map)}\map + \Omega^{-1} m_i\right)  \\
& = \Gamma_i \left(\Omega^{-1}(\map-m_i) + \frac{ \trans{\jacob_{f_i(\map)}}  \jacob_{f_i(\map)}}{\sigma^2}\map + \Omega^{-1}m_i\right) \\
& = \Gamma_i\Gamma_i^{-1}\map = \map\eqs.
\end{align}

\section{Numerical applications}\label{appendix:numerical}
\subsection{A pharmacokinetic example}\label{appendix:numerical.warfa}

\begin{table}[h!]
  \begin{center}
    \caption{MSJD and ESS per dimension.}
    \label{table:msjdandess}
    \begin{tabular}{lccccccr} 
      & &  \textbf{MSJD}& & &  & \textbf{ESS} &  \\
            \hline
       & $ka_i$ & $V_i$ & $k_i$& & $ka_i$ & $V_i$ & $k_i$ \\
      \hline
      RWM & 0.009 & 0.002 & 0.006 & & 1728 & 3414 & 3784   \\
      \textbf{nlme-IMH (Gaussian)}            & 0.061 & 0.004 & 0.018  & &  13694 & 14907 & 19976 \\
	\textbf{nlme-IMH (Student)}            & 0.063 & 0.004 & 0.018  & & 14907 & 19946 & 19856 \\
    \end{tabular}
  \end{center}
\end{table}

\begin{figure}[ht]
  \centering
  \begin{subfigure}{\linewidth}
    \centering
    \includegraphics[width=.8\linewidth]{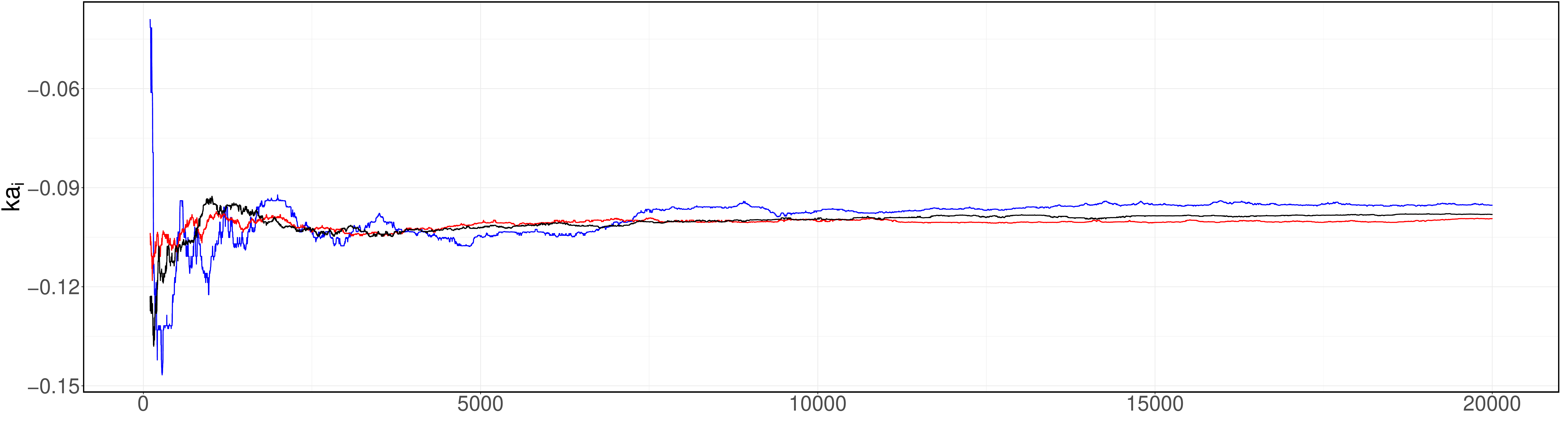}
  \end{subfigure}

  \begin{subfigure}{\linewidth}
    \centering
    \includegraphics[width=.8\linewidth]{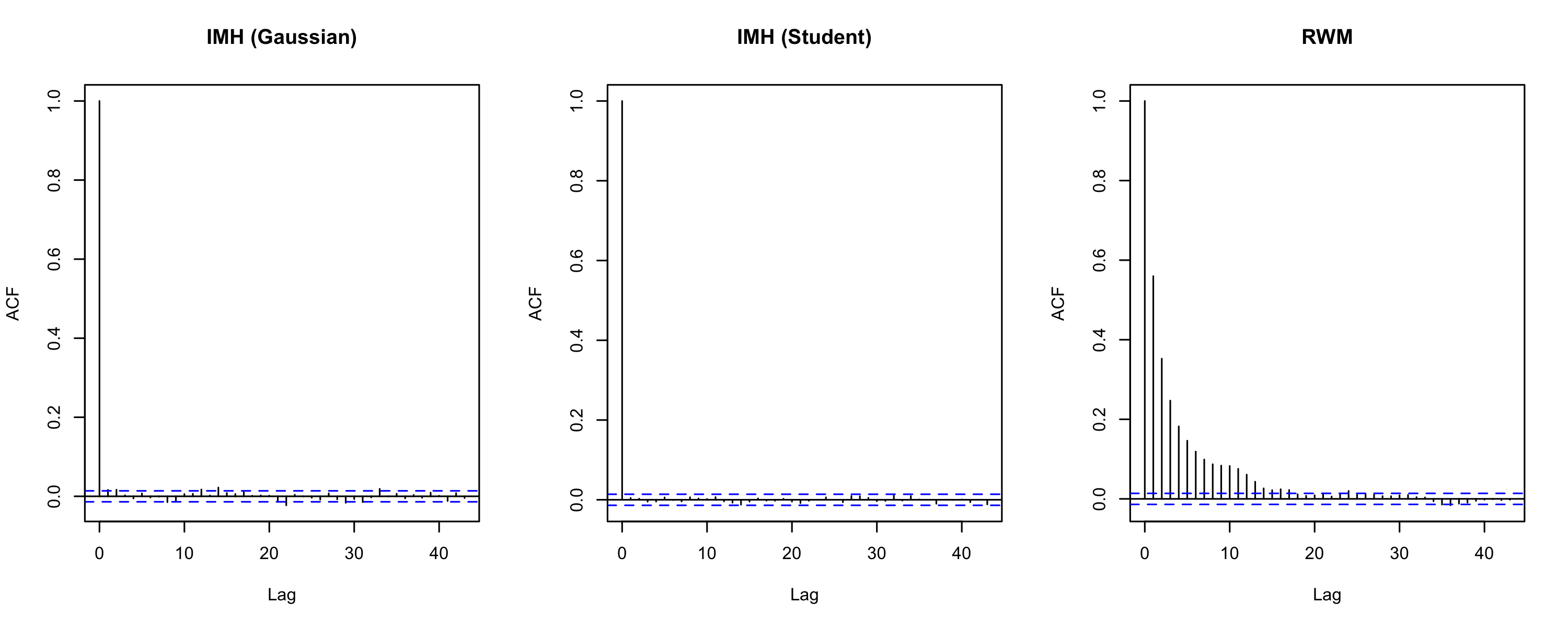}
  \end{subfigure}  
\caption{Modelling of the warfarin PK data. Top plot: convergence of the empirical medians of $\dens(ka_i | y_i ; \theta)$ for a single individual. Comparison between the reference MH algorithm (blue) and the nlme-IMH (red). Bottom plot: Autocorrelation plots of the MCMC samplers for parameter $ka_i$.}
 \label{median_pk_k}
\end{figure}

\begin{figure}[ht]
  \centering
  \begin{subfigure}{\linewidth}
    \centering
    \includegraphics[width=.8\linewidth]{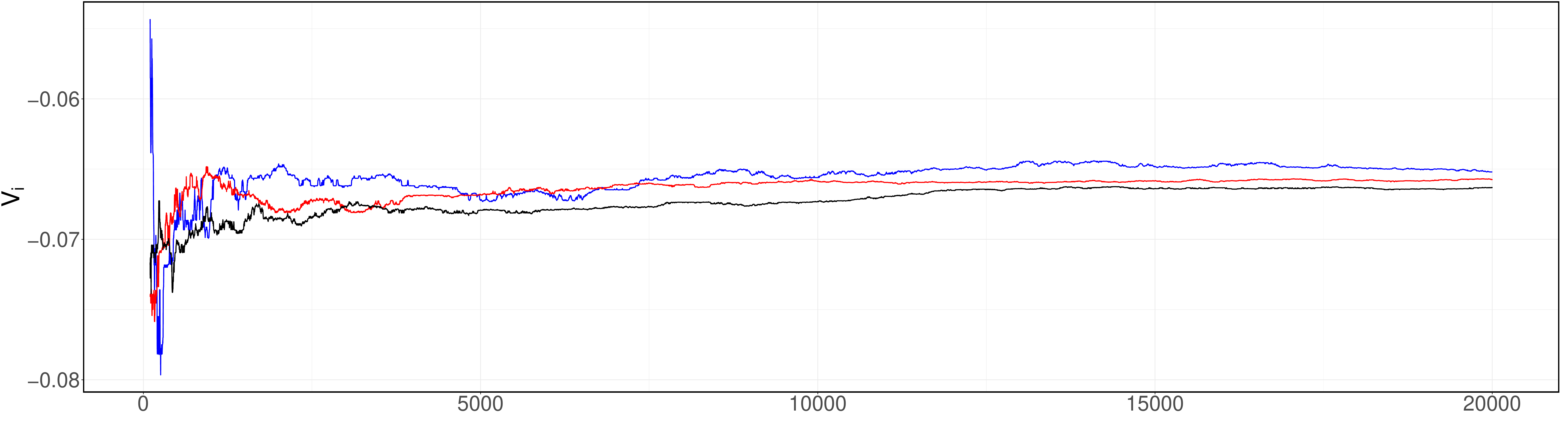}
  \end{subfigure}

  \begin{subfigure}{\linewidth}
    \centering
    \includegraphics[width=.8\linewidth]{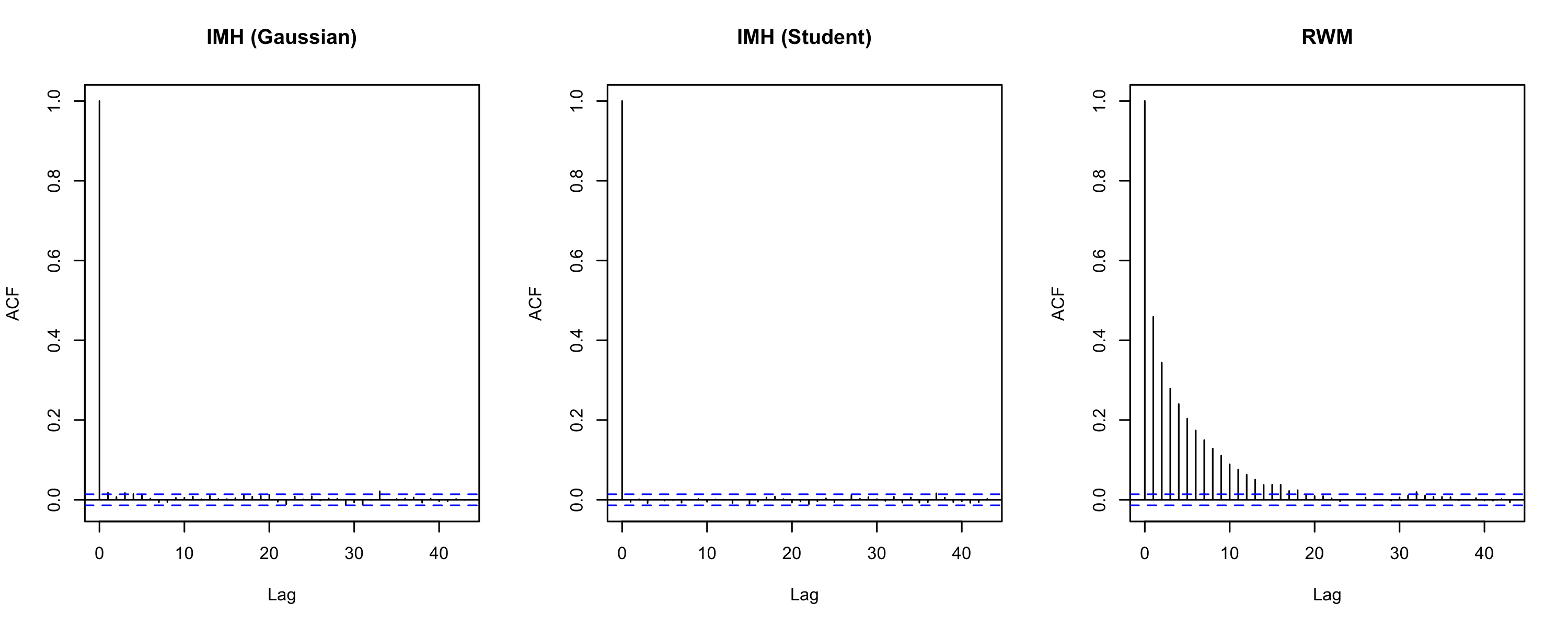}
  \end{subfigure}  
\caption{Modelling of the warfarin PK data. Top plot: convergence of the empirical medians of $\dens(V_i | y_i ; \theta)$ for a single individual. Comparison between the reference MH algorithm (blue) and the nlme-IMH (red). Bottom plot: Autocorrelation plots of the MCMC samplers for parameter $V_i$.}
 \label{median_pk_V}
\end{figure}

Figures~\ref{median_pk_k} and \ref{median_pk_V}  highlight the performances of the RWM, the nlme-IMH using a Gaussian proposal distribution and a Student proposal. At iteration $(t)$ of the MH algorithm, samples from the Student proposal distribution are obtained using the same parameters obtained in Proposal~\ref{lemma:cont} as follows:
\begin{itemize}
\item Student samples $S_i^{(t)}$ are drawn from a student distribution with degree of freedom $k = 3$: $S_i^{(t)} \sim t(k)$
\item Individual parameters $\psi_i^{(t)}$ are obtained using the mean and the covariance defined in Proposal~\ref{lemma:cont} to shift and scale the obtained samples: $\psi_i^{(t)}  = \map^{(t)} + S_i^{(t)}.\Gamma_i^{(t)} $
\end{itemize}

%
%

%
%

\begin{table}[h!]
  \begin{center}
    \caption{Pairwise correlations of the proposals.}
    \label{table:correlation}
    \begin{tabular}{lccr} 
      & $ka_i,V_i$ & $ka_i,k_i$ & $V_i,k_i$ \\
      \hline
      Variational proposal & 0.48 & -0.28 & -0.61   \\
      \textbf{Laplace proposal}          & 0.56 & -0.39 & -0.68  \\
       NUTS (ground truth) & 0.55 &  -0.39 &  -0.68  \\
    \end{tabular}
  \end{center}
\end{table}

\subsection{Time-to-event Data Model}\label{appendix:numerical.tte}
\begin{figure}[ht]
  \centering
  \begin{subfigure}{\linewidth}
    \centering
    \includegraphics[width=.8\linewidth]{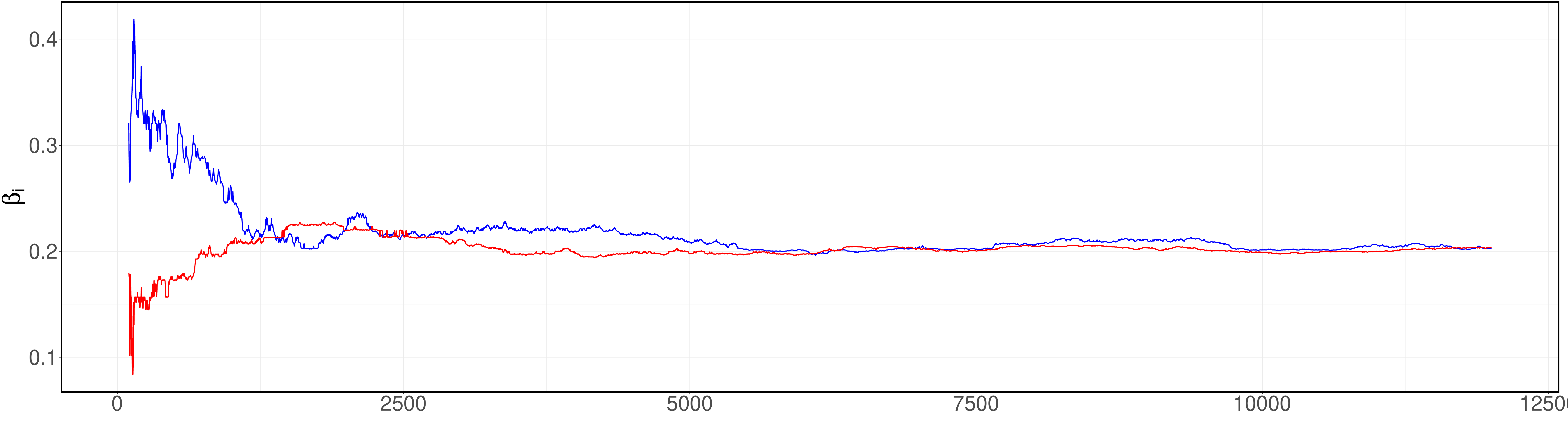}
  \end{subfigure}

  \begin{subfigure}{\linewidth}
    \centering
    \includegraphics[width=.8\linewidth]{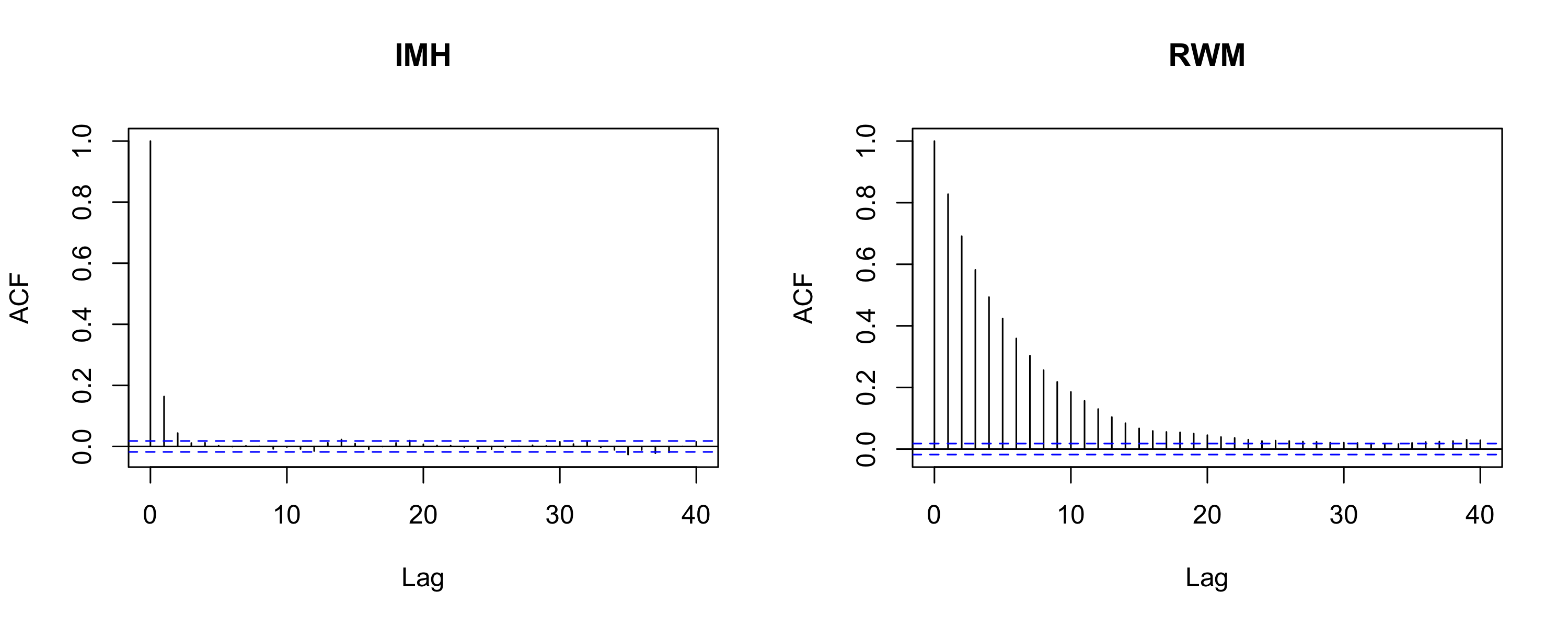}
  \end{subfigure}  
 \caption{Time-to-event data modelling. Top plot: convergence of the empirical medians of $\dens(\beta_i | y_i ; \theta)$ for a single individual. Comparison between the reference MH algorithm (blue) and the nlme-IMH (red). Bottom plot: Autocorrelation plots of the MCMC samplers for parameter $\beta_i$.}
 \label{median_tte_beta}
\end{figure}

Median convergence and autocorrelation plots of the RWM and our nlme-IMH methods for parameter $\beta_i$ are presented in Figure~\ref{median_tte_beta}. Same observations as for parameter $\lambda_i$ can be made.

\end{appendices}

\end{document}